\documentclass[12pt]{iopart}
\usepackage[russian]{babel}
\usepackage{color}

\begin{document}

\title[Balance equations in semi-relativistic quantum hydrodynamics]{Balance equations in semi-relativistic quantum hydrodynamics}

\author{A Yu Ivanov, P A Andreev, L S Kuz'menkov}

\address{Faculty of Physics, Moscow State University, Leninskie Gory build. 1 struct. 2, Moscow, 119991, Russian Federation}
\ead{alexmax1989@mail.ru}

\begin{abstract}
Method of the quantum hydrodynamics has been applied in quantum plasmas studies. As the first step in our consideration, derivation of classical semi-relativistic (i. e. described by the Darwin Lagrangian on microscopic level) hydrodynamical equations is given after a brief review of method development. It provides better distinguishing between classic and quantum semi-relativistic effects. Derivation of the classical equations is interesting since it is made by a natural, but not very widespread method. This derivation contains explicit averaging of the microscopic dynamics. Derivation of corresponding quantum hydrodynamic equations is presented further. Equations are obtained in the five-momentum approximation including the continuity equation, Euler and energy balance equations. It is shown that relativistic corrections lead to presence of new quantum terms in expressions for a force field, a work field etc. The semi-relativistic generalization of the quantum Bohm potential is obtained. Quantum part of the energy current, which is an analog of the quantum Bohm potential for the energy evolution equation, is derived. The Langmuir wave dispersion in semi-relativistic quantum plasmas, corresponding to the Darwin Lagrangian, is also considered to demonstrate contribution of semi-relativistic effects on basic plasma phenomenon.
\end{abstract}


\newcommand{\pia}{\partial_{i\alpha}}
\newcommand{\pib}{\partial_{i\beta}}
\newcommand{\pig}{\partial_{i\gamma}}
\newcommand{\pja}{\partial_{j\alpha}}
\newcommand{\pjb}{\partial_{j\beta}}
\newcommand{\pjg}{\partial_{j\gamma}}

\newcommand{\dia}{D_i^\alpha}
\newcommand{\dib}{D_i^\beta}
\newcommand{\dig}{D_i^\gamma}
\newcommand{\dja}{D_j^\alpha}
\newcommand{\djb}{D_j^\beta}
\newcommand{\djg}{D_j^\gamma}
\newcommand{\dka}{D_k^\alpha}
\newcommand{\dkb}{D_k^\beta}
\newcommand{\dkg}{D_k^\gamma}

\newcommand{\diac}{D_i^{*\alpha}}
\newcommand{\dibc}{D_i^{*\beta}}
\newcommand{\digc}{D_i^{*\gamma}}
\newcommand{\djac}{D_j^{*\alpha}}
\newcommand{\djbc}{D_j^{*\beta}}
\newcommand{\djgc}{D_j^{*\gamma}}
\newcommand{\dkac}{D_k^{*\alpha}}
\newcommand{\dkbc}{D_k^{*\beta}}
\newcommand{\dkgc}{D_k^{*\gamma}}

\newcommand{\dis}{D_{i}^2}
\newcommand{\djs}{D_{j}^2}
\newcommand{\diq}{D_{i}^4}
\newcommand{\djq}{D_{j}^4}
\newcommand{\disc}{D_{i}^{*2}}
\newcommand{\djsc}{D_{j}^{*2}}
\newcommand{\diqc}{D_{i}^{*4}}
\newcommand{\djqc}{D_{j}^{*4}}

\newcommand{\via}{v_{i\alpha}}
\newcommand{\vib}{v_{i\beta}}
\newcommand{\vig}{v_{i\gamma}}
\newcommand{\vja}{v_{j\alpha}}
\newcommand{\vjb}{v_{j\beta}}
\newcommand{\vjg}{v_{j\gamma}}
\newcommand{\uia}{u_{i\alpha}}
\newcommand{\uib}{u_{i\beta}}
\newcommand{\uig}{u_{i\gamma}}
\newcommand{\uja}{u_{j\alpha}}
\newcommand{\ujb}{u_{j\beta}}
\newcommand{\ujg}{u_{j\gamma}}

\section{Introduction}

Paper is dedicated to further development of the many-particle quantum hydrodynamics \cite{LSK1999}-\cite{arxiv3335}. In previous works on this subject non-relativistic systems of particles and also influence of several (particularly, spin related) relativistic corrections in Hamiltonian describing system have been studied in \cite{LSK2001}, \cite{arxiv3335}, \cite{Andreev RPJ2007}. Main goal of this work is an obtaining of the quantum hydrodynamical equations for system of charged particles with the Coulomb and the current-current interactions, placed into an external electromagnetic field. The relativistic correction to the kinetic energy of particles is also taken into consideration. Historically the Biot-Savart-Laplace law was among the first examples of the current-current interaction. It is possible to formulate the problem with electromagnetic interaction of N particles. Lagrangian of this system in semi-relativistic approximation to the second order of $v/c$ was derived by Darwin \cite{LaLi2} and has form
\begin{eqnarray}\label{Darwin_Lagr}
&&L=\sum_{i=1}^N \biggl(\frac{m_i v_i^2}{2}+\frac{m_i v_i^4}{8c^2}\biggr)-\frac{1}{2}\sum_{i,j=1,i\neq j}^N \frac{e_i e_j}{r_{ij}}\nonumber\\
&&{}+\sum_{i,j=1,i\neq j}^N \frac{e_i e_j}{4c^2}\biggl[\frac{\mathbf{v}_i\mathbf{v}_j}{r_{ij}}+\frac{(\mathbf{v}_i\mathbf{r}_{ij})
(\mathbf{v}_j\mathbf{r}_{ij})}{r_{ij}^3}\biggr].
\end{eqnarray}
where $m_i$, $e_i$ are masses and charges of particles, $c$ is the speed of light, $\mathbf{r}_i$ is radius vector of $i$-th particle, $\mathbf{v}_i=\dot\mathbf{r}_i$ are velocities of particles, $\mathbf{r}_{ij}=\mathbf{r}_{i}-\mathbf{r}_{j}$. The first group of terms in the first line is the non-relativistic kinetic energy of particles, the second group is associated with the relativistic correction to it, the third is energy of the Coulomb interaction. Group of terms in the second line correspond to the current-current interaction. Radiation of electromagnetic waves is not included into Lagrangian in this approximation.

Kinetic equation describing dynamics of N charged interacting particles in self-consistent field approximation was proposed by A. A. Vlasov \cite{Vlasov UFN}, \cite{Bohm Gross} in 1938. In the process full electromagnetic interaction was taken into account, i. e. fields produced by particles satisfy to the full set of Maxwell equations. Derivation of a kinetic equation for N particle system with the Coulomb interaction was presented in 1946 by N. N. Bogolyubov, it is known as the BBGKY hierarchy (the method can be found in \cite{LaLi10}, \cite{Klimontovich book}). Derivation of the kinetic equation based on the Darwin Lagrangian presented in \cite{Zaslavskii 62}, \cite{Pavlotskii DAN 73}, \cite{Pavlotskiy}, some further development of this field can be found in Refs. \cite{Pavlotsky P A 89 2}-\cite{Mingalev PL A 96}. Consideration of the semi-relativistic effects in quantum plasmas was started in Refs. \cite{Goldstein Ph 77}-\cite{Jones PF 80 2}, some of it was done in terms of the Wigner quantum distribution function \cite{Wigner PR 84}. Recent papers show that there is strong interest to relativistic effects in plasmas \cite{Shah PP 11}-\cite{Mahajan Yoshida} and quantum plasmas (see for instance Ref. \cite{Asenjo ZMBJ}). The semi-relativistic corrections be of interest also in equations of quantum plasma kinetics \cite{Asenjo ZMBJ}. The spin-orbit interaction in quantum plasmas was considered in \cite{arxiv3335}, \cite{Andreev DSS 09}-\cite{Trukhanova}. Quantum hydrodynamics based on the relativistic Klein-Gordon equations was considered in \cite{Haas Eliasson Shukla PRE2012}. Quantum hydrodynamical description of the Dirac electron was presented in \cite{Asenjo Munoz PhysPlasm}. Many astrophysical applications wait for creation of relativistic quantum hydrodynamics and kinetics. Some of these astrophysical phenomena were described in recent review \cite{Uzdensky arxiv review 14}.

Some properties of the semi-relativistic many-particle QHD have been considered in our recent papers \cite{Ivanov arxiv 12} and \cite{Ivanov RPJ 13}. These research were focused on the Euler equation and Langmuir wave dispersion. The Darwin interaction was also considered in \cite{Ivanov arxiv 12}. It was shown that the Darwin interaction competes with contribution given by the semi-relativistic part of the kinetic energy.

Creation of fully relativistic quantum hydrodynamics meets some complications, because quantum equation describing dynamics of N relativistic particles with full electromagnetic interaction does not exist. This system of particles is non-Hamiltonian system. Evolution of one particle with spin in external electromagnetic field is described by the Dirac equation.

Quantum generalization of the Hamiltonian corresponding to the Darwin Lagrangian was proposed by Breit \cite{Breit}. So, on the way of relativistic quantum hydrodynamics construction is naturally to use this Hamiltonian describing quantum system of N particles with electromagnetic interaction to the second order of $v/c$. Quantum mechanical derivation of the Breit Hamiltonian was done by L. D. Landau; derivation from scattering amplitude of two electrons is presented in Ref. \cite{LaLi4} (see section 83). The Breit Hamiltonian has form
\begin{eqnarray}
&&\hat H=\sum_{i=1}^N\biggl(\frac{\hat\mathbf{p}_i^2}{2m_i}-\frac{\hat \mathbf{p}^4_i}{8m^3_ic^2}\biggr)\nonumber\\
&&{}+\frac{1}{2}\sum_{i,j=1,i\neq j}^N\biggl\{\frac{e_i e_j}{r_{ij}}-\pi\frac{e_i e_j\hbar^2}{2c^2}\biggr(\frac{1}{m_i^2}+\frac{1}{m_j^2}\biggl)\delta(\mathbf{r}_i-\mathbf{r}_j) \nonumber\\
&&{}
-\frac{e_i e_j}{2m_i m_j c^2r_{ij}}\biggl(\hat \mathbf{p}_i\hat\mathbf{p}_j+\frac{\mathbf{r}_{ij}(\mathbf{r}_{ij}\hat\mathbf{p}_i)
\hat\mathbf{p}_j}{r_{ij}^2}\biggr)\nonumber\\
&&{}-\frac{e_i e_j\hbar}{4c^2r_{ij}^3}\biggl[\frac{[\mathbf{r}_{ij}\hat\mathbf{p}_i]
_{\alpha}\sigma_{i\alpha}}{m_i^2}-\frac{[\mathbf{r}_{ij}\hat\mathbf{p}_j]
_{\alpha}\sigma_{j\alpha}}{m_j^2}+\frac{2([\mathbf{r}_{ij}\hat\mathbf{p}_i]
_{\alpha}\sigma_{j\alpha}-[\mathbf{r}_{ij}\hat\mathbf{p}_j]
_{\alpha}\sigma_{i\alpha})}{m_i m_j}\biggr]\nonumber\\
&&{}+\frac{e_i e_j\hbar^2}{8m_i m_j c^2}\sigma_{i\alpha}\sigma_{j\beta}\biggl[\frac{\delta^{\alpha\beta}}{r_{ij}^3}-
\frac{3x_{\alpha ij}x_{\beta ij}}{r_{ij}^5}-\frac{8\pi}{3}\delta^{\alpha\beta}
\delta(\mathbf{r}_i-\mathbf{r}_j)\biggr]\biggr\},
\end{eqnarray}
where $\hat\mathbf{p}_i=-i\hbar\mathbf{\nabla}_i$ is momentum operator of particle, $\sigma_{i\alpha}$ is spin operator of particle, $\hbar$ is the Planck constant. As it is seen from this expression, the Breit Hamiltonian contains, along with the Coulomb and the current-current interactions, also the spin-spin (the fourth line), spin-current and spin-orbit interactions (the third line). There is also interaction proportional to $\delta(\mathbf{r}_i-\mathbf{r}_j)$, which corresponds to the Zitterbewegung effect.

Through studies of specific physical processes, such as dispersion of waves, interaction of charged beams with plasmas, equations of quantum hydrodynamics have been used in modern literature, and these equations can be derived from both the many-particle Schr\"{o}dinger equation to be truncated for particular system of particles \cite{LSK1999}, \cite{LSK2001}, \cite{Andreev LSK Trukhanova PRB}, \cite{Andreev LSK 2008} \textit{and} from the Schr\"{o}dinger equation for one particle in an external field \cite{Haas}-\cite{Brodin Marklund 2007}.

Since we develop the method of quantum hydrodynamics, let us briefly mention history of its development. In 1926 Madelung \cite{Madelung}, on account of the Schr\"{o}dinger equation for one particle in an external field, derived a set of equation in hydrodynamical form. This set of equations consists from the continuity equation and the momentum balance equation. Velocity field is potential there. In this way, Madelung made a transfer from one abstract complex function to the observable physical quantities - probability density and velocity field. Hydrodynamical formulation of quantum problem on motion of charged particle with a spin in an external electromagnetic field was accomplished by Takabayashi in 1955 \cite{Takabayashi}. In his work, along with the continuity and momentum balance equations, an equation of spin evolution was obtained. An explicit form of the force field associated with magnetic moment interaction with an external field was derived in the momentum balance equation. So hydrodynamics may appear directly from equations of mechanics. Nevertheless, the traditional way of derivation of hydrodynamic equation is getting of moments of kinetic equation. To the end one can use the Wigner kinetic equation \cite{Klimontovich book}, \cite{Jones 78 1}, \cite{Wigner PR 84}, \cite{Haas PRE 00}. However, it is possible to find straightforward derivation of quantum hydrodynamics from many-particle Schr\"{o}dinger equation. It was done in 1999-2001 in Refs.\cite{LSK1999}, \cite{LSK2001}, \cite{MaksimovTMP 2001 b}. Further development was made in Refs. \cite{Andreev LSK Trukhanova PRB}, \cite{Andreev RPJ2007}, \cite{Ivanov RPJ 13}. Quantum hydrodynamics has different applications, but most of all it has been used for studying of quantum plasmas. Reviews of recent applications of the quantum hydrodynamics for quantum plasmas were presented in Refs. \cite{Shukla Eliasson 2010 rus}, \cite{Shukla Eliasson 2011}.

In Ref. \cite{Kuzelev Ruhadze UFN1999} effectiveness of hydrodynamical description
(obtaining from the one-particle Schr\"{o}dinger equation) for kinetic properties
of quantum plasmas was demonstrated.

Investigating quantum plasma of many-electron atom, many-nucleon problem, transport phenomena, propagation modes we have to deal with dynamics of many interacting particles. Schr\"{o}dinger equation for a system of N interacting particles is defined in a 3N dimensional configuration space, when propagation of excitations, processes of momentum and energy exchange are taken place in three dimensional physical space. So, there is problem of conversion from function defined in physical space to an equivalent quantum description of system in terms of physical fields in three dimensional space. The method of many-particle quantum hydrodynamics \cite{LSK1999}, \cite{LSK2001}, \cite{Andreev LSK Trukhanova PRB} solves this problem allowing to get dynamics in physical space. This transition is reached by the Dirac delta function existing in definition of hydrodynamic variables. They play role of projection operators from 3N dimensional configurational space to three dimensional physical space.

In the first paper on \cite{LSK1999} quantum hydrodynamic method for N interacting particles the Coulomb interaction only is taken into account. Many-particle QHD allows to derive the continuity and Euler equations, but it also allows to derive other hydrodynamic equations, such as the energy, pressure, energy flux evolution equations, and make truncation at necessary point. The energy evolution equation were derived in Ref. \cite{LSK1999}. Quantum contributions in the momentum and energy balance equations were obtained. Quantum exchange correlations for bosons and fermions were considered in \cite{LSK1999} as well. It was done to consider interparticle interaction beyond the self-consistent field approximation.

In following papers other interactions were included in the QHD scheme: the spin-spin interaction \cite{LSK2001}, the spin-current interaction \cite{arxiv3335}, \cite{Andreev RPJ2007}, and the spin-orbit interaction \cite{arxiv3335}, \cite{Andreev DSS 09}, \cite{pavelproc}. Exchange part of the Coulomb and spin-spin interactions were derived in Ref. \cite{MaksimovTMP 2001 b}. It was applied for calculation of wave dispersion in quantum plasmas \cite{Andreev AtPhys 08}. In other papers, the ultra-cold quantum gases of neutral particles \cite{Andreev LSK 2008}, systems of particles with the electric dipole-dipole interactions, where equations of polarization evolution were derived and applied \cite{Andreev LSK Trukhanova PRB}, \cite{Andreev LSK arxiv12}, system of spinning particles with electric dipole interactions \cite{Trukhanova} etc were investigated. We should mention that QHD theory of the ultra-cold quantum gases of neutral particles is based upon the exchange part of the short-range interaction \cite{Andreev LSK 2008}.

Evolution of magnetic moments in quantum plasmas leads to different phenomena \cite{Shukla Eliasson 2010 rus}. Particularly it supports propagation of new type of excitations \cite{Andreev VestnMSU 2007}, \cite{Brodin PRL 08}, \cite{Vagin 09}, where spin dynamics are involved in evolution of electromagnetic perturbations. So we can call them spin-electromagnetic plasma waves. Existence of specific spin waves, where the electric field existing in plasma gives no contribution in mechanism of propagation of these spin waves, was also shown in Ref. \cite{Andreev VestnMSU 2007}. More detailed description of the spin waves and spin-electromagnetic plasma waves can be found in Ref. \cite{arxiv3335}. Instabilities of magnetized quantum plasmas can be caused by propagation of neutron beam \cite{arxiv3335}, \cite{Andreev AtPhys 08}. These instabilities are caused by the spin-spin interaction of neutron spins and plasma spins, \emph{and} also the spin-current interaction of neutron beam spins and electric currents in plasma.

Some papers dedicated to discussion of applicability area for the QHD have been appeared \cite{Vladimirov PU 11}, \cite{Manfredi FIC 05}. We should mention that these debates are conducted around the QHD equations obtained from the one-particle Schr\"{o}dinger equation or the many-particle Schr\"{o}dinger equation splitted on number of one-particle equations for particles with weak interaction.

Interactions consisting the Breit Hamiltonian enter in it in additive way, and, consequently, results obtained for every interaction separately give full result by summation of each other. As next step in investigations of quantum hydrodynamics developed on base of the Breit Hamiltonian, in this work the current-current interaction and the relativistic correction to kinetic energy of particle are considered. It gives opportunity to use derived framework for wide class of physical systems, considering, for example, that solar plasma is not ultrarelativistic ($T\ll mc^2$). The current-current interaction plays significant role at investigations of charged beams.

Application of the Wigner kinetics spreads with increases of interest to quantum plasmas in the last years \cite{Haas PRE 00}, \cite{Manfredi PRB 01}-\cite{Haas NJP 10}, including generalizations for spinning particles. In most cases authors use the Wigner distribution function defined in terms of one-particle wave functions of independent particles obeying the one-particle Schr\"{o}dinger equations. As the result it gives closed mathematical apparatus looking like quantum Vlasov equation in the self-consistent field approximation. However, the general Wigner distribution function may be used for derivation of more general chain of kinetic equation to be truncated at necessary step \cite{Wigner PR 84}. The Wigner kinetic technic was also applied in Refs. \cite{Trovato JP A 10}, \cite{Trovato PRE 10}, where authors consider interaction under some general condition and derive corresponding quantum hydrodynamic equations applying the principle of quantum maximum entropy to close set of equations. Some attempts to get alternative kinetic methods have been also recently performed \cite{Altaisky PL A 10}-\cite{Andreev kinetics 13}.

Having kinetic equation one can derive corresponding quantum hydrodynamic equations, so that obtained set does not restricted by the continuity and Euler equations (corresponding set of QHD equations for spinning particles also contains the generalized Bloch equation for the spin density evolution). It allows to find equations for the energy evolution, the pressure evolution etc. We have not tested capabilities of the Wigner kinetics, but we have not seen derivation of explicit form of the momentum and energy fluxes containing quantum Bohm potentials. Method of the many-particle QHD, which is under derivation in this paper, also allows to get equations for evolution of the energy, pressure, energy flux etc, but in opposite to the Wigner kinetics, it allows to obtain explicit form of the fluxes including quantum parts related to the quantum Bohm potential (see for instance \cite{LSK1999}, \cite{LSK2001}, \cite{Andreev RPJ2007}, and below in this paper). Non-relativistic quantum energy evolution equation including the Bohm potential was considered in Ref. \cite{Ivanov arxiv 13} to get influence of the energy evolution on the Langmuir wave dispersion.

One more method of QHD derivation has been suggested recently \cite{Koide PRC 13}. This method also gives full chain of hydrodynamic equations. Particularly, in Ref. \cite{Koide PRC 13} author discuss derivation the energy density, the positivity of the entropy production
is used in the derivation \emph{and} hence the viscous tensors
and the heat current are obtained applying
the linear irreversible thermodynamics to construct a
hydrodynamic model with spin. The method suggested by T. Koide allows to obtain full chain of hydrodynamic equation for quantum mediums as we do it in the many-particle QHD. However, we do not apply concepts additional to quantum mechanics and keep to stand in terms of one concept. Nevertheless, irreversibility is a fundamental property of processes in nature, and its including enriches quantum theory.

This paper is dedicated to wider consideration of the many-particle QHD method for charged spinless particles in the semi-relativistic approximation. Comparison with the classic approach and account of the energy evolution were also performed in the paper.

This paper is organized as follows. In Section 2 we derive equations for the classic semi-relativistic hydrodynamics of plasmas. In Section 3 description of QHD method is given. In Sections 4-6 we derive continuity equation, momentum and energy balance equations correspondingly. Velocity field is introduced, and obtained equations are expressed through it in Section 7. Semi-relativistic part of the quantum Bohm potential is derived in Section 7. An explicit form of the quantum thermal current up to the semi-relativistic contribution is also obtained in Section 7. As an example, in Section 8 we discuss the longitudinal waves in semi-relativistic plasmas. Conclusion is presented in Section 9.

\section{Foundations of classical semi-relativistic hydrodynamics}\label{classic_hd}

Until we analyze equation of quantum hydrodynamics, let us deduce the classical equations. It needs to be done for comparison with quantum equations obtained later. This consideration is conducted with method suggested in 1996 \cite{Drofa1996}. It was a branch of more general scheme \cite{LSK 91}. Recent discussion of some method details can be found in Ref. \cite{pavelproc cl}. This method allows to obtain hydrodynamical equations from microscopic equations of dynamics for N particles and definition of microscopic density. Some properties of relativistic hydrodynamic was considered in terms of this method in Ref. \cite{Andreev arxiv rel}.

Using Lagrange equations
\begin{equation}\frac{d}{dt}\frac{\partial L}{\partial v_{i\alpha}}=\frac{\partial L}{\partial x_{i\alpha}},\end{equation}
we obtain classical equation of motion derived from the Darwin Lagrangian (\ref{Darwin_Lagr}). From this we express acceleration of particle $\dot v_{i\alpha}$ inverting structure placed before it. It cannot be carried out exactly, but, keeping terms only to the second order of $v/c$, we obtain next result:
\begin{eqnarray} \label{acceleration}
\dot v_{i\alpha}&=&\frac{e_i}{m_i}\biggl[\biggl(1-\frac{v_i^2}{2c^2}\biggr)\delta_{\alpha\beta}-
\frac{1}{c^2}v_{i\alpha}v_{i\beta}\biggr](E_{i\beta}^{ext}+E_{i\beta}^{int})\nonumber\\
&&{}+\frac{e_i}{m_i c}\varepsilon_{\alpha\beta\gamma}v_{i\beta}(B_{i\gamma}^{ext}+B_{i\gamma}^{int})
+\sum_{j=1,j\neq i}^N\frac{e_i e_j}{2m_i c^2}v_{j\beta}v_{j\gamma}\partial_{i\beta}G^{\alpha\gamma}_{ij}\nonumber\\
&&{}-\sum_{j=1,j\neq i}^N\frac{e_i e_j^2}{2m_i m_j c^2}G^{\alpha\beta}_{ij}(E_{j\beta}^{ext}+E_{j\beta}^{int}),
\end{eqnarray}
where field $E^{int}_\alpha$ contains only potential part.

In correspondence with \cite{Drofa1996}, \cite{pavelproc cl} we define mass density in the adjacency $\Delta(\mathbf{r})$ of physically infinitesimal volume with value $\Delta$ in following way:
\begin{equation}\rho(\mathbf{r},t)=\frac{1}{\Delta}\int_{\Delta(\mathbf{r})}d\xi\sum_{i=1}^N m_i\delta(\mathbf{r}+\xi-\mathbf{r}_i(t)).\end{equation}
For particles of one species, for instance electrons, we have simple relation between the mass density $\rho$ and the particle concentration $n$: $\rho(\mathbf{r},t)=mn(\mathbf{r},t)$.

Differentiating this expression with respect to time $t$, the continuity equation is obtained:
\begin{equation}\label{cont eq classicic}\partial_t \rho(\mathbf{r},t)+\partial_\alpha j^\alpha(\mathbf{r},t)=0,\end{equation}
where current density is
\begin{equation}j^\alpha(\mathbf{r},t)=\frac{1}{\Delta}\int_{\Delta(\mathbf{r})}d\xi\sum_{i=1}^N m_i v_{i\alpha}(t)\delta(\mathbf{r}+\xi-\mathbf{r}_i(t)).\end{equation}

Differentiating with respect to time the particle current, an
equation of the current evolution is obtained. This equation
corresponds to the Euler equation in hydrodynamics:
\begin{equation}\label{classic momentum bal eq}\partial_t j^\alpha+\partial_\beta\Pi^{\alpha\beta}=en E^\alpha+\frac{e}{mc}\ \varepsilon^{\alpha\beta\gamma}j^\beta
B^\gamma+\mathcal{F}_\alpha,\end{equation}
where we have used $\rho(\mathbf{r},t)=mn(\mathbf{r},t)$.
For simplicity we consider system with one kind of particles, but we keep in mind that plasmas consist of two or more kinds of
particles. If we would consider several kinds of particles, we have a set of QHD equations for each kind of particles. All these equations are coupled via electromagnetic field created by particles and satisfying to the Maxwell equations. The first two terms in the
right-hand side represent the Lorentz force and describe an action
of an electromagnetic field on charges and currents. The first
term in the right-hand side of equation (\ref{classic momentum bal
eq}) represents sum of the external electric field and the Coulomb
interaction. The second term describes an action of the external
magnetic field on the currents of system, and also the
current-current interaction. $\mathcal{F}_\alpha$ is density of
force caused by the semi-relativistic effects, namely by the
relativistic correction to the kinetic energy and the
current-current interaction. The current-current interaction
enters both the Lorentz force and $\mathcal{F}_\alpha$. Let us
introduce a notion of thermal velocity $\uia$ particle as
difference between the velocity of particle and the velocity
field:
$u_{i\alpha}(\mathbf{r},t)=v_{i\alpha}(t)-v_\alpha(\mathbf{r},t)$,
where the velocity field $v_\alpha$ is defined by formula
$j_\alpha(\mathbf{r},t)=mn(\mathbf{r},t)v_\alpha(\mathbf{r},t)$.
The current on thermal velocities equals to zero, thereby velocity
field $v_\alpha$ is separated. The flux of
particle current equals to
\begin{equation}\Pi_{\alpha\beta}(\mathbf{r},t)=mnv_\alpha v_\beta(\mathbf{r},t)+p_{\alpha\beta}(\mathbf{r},t),\end{equation}
where $p_{\alpha\beta}$ is the tensor of kinetic pressure having
form
\begin{equation}p_{\alpha\beta}(\mathbf{r},t)=\frac{1}{\Delta}\int_{\Delta(\mathbf{r})}
d\xi\sum_{i=1}^N m_i u_{i\alpha}u_{i\beta}\delta(\mathbf{r}+\xi-\mathbf{r}_i(t)).\end{equation}
The electric and magnetic fields represent the sum of external
fields and fields caused by charges of particles:
$E^\alpha=E^\alpha_{ext}+E^\alpha_{int}$, $B^\alpha=B^\alpha_{ext}+B^\alpha_{int}$.

The force density $\mathcal{F}_\alpha$ take the form
\begin{eqnarray}
\label{fla5}
&&\mathcal{F}_\alpha=-\frac{e}{mc^2}\biggl[\delta_{\alpha\beta}\biggl(\frac{1}{2}mnv^2+
\rho\epsilon\biggr)+(mnv_\alpha v_\beta+p_{\alpha\beta})\biggr]E_\beta\nonumber\\
&&{}+\frac{e^2}{2c^2}\int d\mathbf{r}'[\partial_\alpha G_{\beta\gamma}(\mathbf{r}-\mathbf{r}')-\partial_\beta G_{\alpha\gamma}(\mathbf{r}-\mathbf{r}')]\pi_{\beta\gamma}(\mathbf{r},\mathbf{r}',t)\nonumber\\
&&{}+\frac{e^2}{2mc^2}n\int d\mathbf{r}'\partial_\gamma G_{\alpha\beta}(\mathbf{r}-\mathbf{r}')[mn(\mathbf{r}',t)v_\beta(\mathbf{r}',t) v_\gamma(\mathbf{r}',t)+p_{\beta\gamma}(\mathbf{r}',t)]\nonumber\\
&&{}-\frac{e^3}{2mc^2}n\int d\mathbf{r}'G_{\alpha\beta}(\mathbf{r}-\mathbf{r}')E_\beta(\mathbf{r}',t)n(\mathbf{r}',t),
\end{eqnarray}
where
\begin{equation}\rho\epsilon(\mathbf{r},t)=\frac{1}{\Delta}\int_{\Delta(\mathbf{r})}
d\xi\sum_{i=1}^N \frac{1}{2}m_i u_{i}^2\delta(\mathbf{r}+\xi-\mathbf{r}_i(t)),\end{equation}
\begin{equation}\pi_{\alpha\beta}(\mathbf{r},\mathbf{r}',t)=\frac{1}{\Delta}
\int_{\Delta(\mathbf{r})}d\xi\sum_{i,j=1,j\neq i}^N
\delta(\mathbf{r}+\xi-\mathbf{r}_i)\delta(\mathbf{r}'-\mathbf{r}_j)u_{i\alpha}u_{j\beta}.
\end{equation}
In formula (\ref{fla5}) terms in the first line arise from the semi-relativistic correction to energy of particle. Function
$\rho\epsilon(\mathbf{r},t)$ represents non-relativistic part of
kinetic energy of thermal motion, and presence of term with tensor
$\pi_{\alpha\beta}(\mathbf{r},\mathbf{r}',t)$ in equations
associated with the fact that thermal motion of particles also
causes magnetic field, which give an influence on evolution of
particle current. The term in the third line appears from the current-current interaction between particles, and the term in the fourth line is caused by dependence of particle acceleration (\ref{acceleration}) from the function of current-current interaction.

Let us consider the energy balance equation in classical
hydrodynamics. This equation is the fifth equation in the five
momentum approximation. Energy density of system is defined by
formula
\begin{eqnarray}
\label{fla7}
\varepsilon(\mathbf{r},t)&=&\frac{1}{\Delta}\int_{\Delta(\mathbf{r})}d\xi\sum_{i=1}^N
\delta(\mathbf{r}+\xi-\mathbf{r}_i)\biggl[\frac{1}{2}m_i v_i^2+\frac{3}{8c^2}m_i v_i^4\nonumber\\
&&{}+\sum_{j\neq i}^N\biggl(\frac{1}{2}e_i e_j G_{ij}+\frac{e_i e_j}{4c^2}
v_{i\alpha}v_{j\beta}G^{\alpha\beta}_{ij}\biggr)\biggr].
\end{eqnarray}
The first group of terms in this formula is density of non-relativistic kinetic energy, the second group is semi-relativistic correction to it, the third and the fourth groups are densities of the Coulomb and the current-current interactions correspondingly.

Applying the procedure described above, we obtain following
equation for the energy density:
\begin{eqnarray}
\label{fla16}
&&\partial_t \varepsilon(\mathbf{r},t)+\partial_\alpha Q^\alpha(\mathbf{r},t)=en(\mathbf{r},t)\mathbf{v}(\mathbf{r},t)\mathbf{E}
+\alpha(\mathbf{r},t)\nonumber\\
&&{}+\frac{e^2}{2}n(\mathbf{r},t)\int d\mathbf{r}'\partial^\alpha G(\mathbf{r}-\mathbf{r}')(v^\alpha(\mathbf{r},t)-v^\alpha(\mathbf{r}',t))n(\mathbf{r}',t)
\nonumber\\
&&{}+\frac{e^3}{4mc^2}n E_\alpha(\mathbf{r},t)\int d\mathbf{r}'G^{\alpha\beta}(\mathbf{r}-\mathbf{r}')nv_\beta(\mathbf{r}',t)\nonumber\\
&&{}-\frac{e^3}{4mc^2}
nv^\alpha(\mathbf{r},t)\int d\mathbf{r}'E^\beta(\mathbf{r}',t)G^{\alpha\beta}(\mathbf{r}-\mathbf{r}')n(\mathbf{r}',t)
\nonumber\\
&&{}+\frac{e^2}{4c^2}nv^\alpha(\mathbf{r},t)\int d\mathbf{r}'\partial^\gamma G^{\alpha\beta}(\mathbf{r}-\mathbf{r}')(v^\gamma(\mathbf{r},t)+v^\gamma(\mathbf{r}',t))
nv^\beta(\mathbf{r}',t),
\end{eqnarray}
where energy density can be represented as
\begin{eqnarray}
&&\varepsilon(\mathbf{r},t)=\frac{1}{2}mnv^2+\frac{e^2}{2}n\int d\mathbf{r}'G(\mathbf{r}-\mathbf{r}')n(\mathbf{r}',t)+\frac{3}{8c^2}mnv^4\nonumber\\
&&{}+\frac{e^2}{4c^2}nv_\alpha\int d\mathbf{r}'G^{\alpha\beta}(\mathbf{r}-\mathbf{r}')n(\mathbf{r}',t)v_{\beta}(\mathbf{r}',t)+
\rho\epsilon,
\end{eqnarray}
function $\rho\epsilon$ is introduced here, which has sense of internal energy density (coinciding with it in non-relativistic theory; in semi-relativistic theory in $\rho\epsilon$ terms combined with velocity field and thermal velocities appear):
\begin{eqnarray}\label{thermal_energy_density}
&&\rho\epsilon(\mathbf{r},t)=\frac{1}{\Delta}\int_{\Delta(\mathbf{r})}d\xi\sum_{i=1}^N \delta(\mathbf{r}+\xi-\mathbf{r}_i(t))\biggl[\frac{1}{2}m_i u_{i}^2\nonumber\\
&&{}+\frac{3m_i}{8c^2}\{4(\mathbf{v}(\mathbf{r},t)\mathbf{u}_i)^2+u_i^4+2v^2(\mathbf{r},t)u_i^2+
4(\mathbf{v}(\mathbf{r},t)\mathbf{u}_i)u_i^2\}\nonumber\\
&&{}+\sum_{j\neq i}^N\frac{e_i e_j}{4 c^2}G^{\alpha\beta}_{ij}\biggl(v_{\alpha}(\mathbf{r},t)u_{j\beta}+u_{i\alpha}
v_{\beta}(\mathbf{r}',t)+u_{i\alpha}u_{j\beta}\biggr)\biggr].
\end{eqnarray}
In formula (\ref{thermal_energy_density}) terms caused by semi-relativistic part of the Lagrangian (\ref{Darwin_Lagr}) are seen again. Terms in the second line appear from relativistic correction to kinetic energy, in the third line - from the current-current interaction.

Function of energy current $Q^\alpha(\mathbf{r},t)$, which some times is called the energy flux, can be presented as a sum of three terms:
\begin{eqnarray}
Q^\alpha(\mathbf{r},t)=v^\alpha\varepsilon+v^\beta p^{\alpha\beta}+q^\alpha,
\end{eqnarray}
where the thermal energy current $q^\alpha(\mathbf{r},t)$ equals
to
\begin{eqnarray}
&&q^\alpha(\mathbf{r},t)=\frac{1}{\Delta}\int_{\Delta(\mathbf{r})}d\xi\sum_{i=1}^N \delta(\mathbf{r}+\xi-\mathbf{r}_i(t))\biggl[\frac{1}{2}mu_{i\alpha}u_i^2+\sum_{j\neq i}^N\frac{1}{2}u_{i\alpha}e_i e_j G_{ij}\nonumber\\
&&{}+\frac{3m}{8c^2}u_{i\alpha}\{4v^2(\mathbf{r},t)v_{\beta}(\mathbf{r},t)u_{i\beta}+
4(\mathbf{v}(\mathbf{r},t)\mathbf{u}_i)^2\nonumber\\
&&{}+u_i^4+2v^2(\mathbf{r},t)u_i^2+
4(\mathbf{v}(\mathbf{r},t)\mathbf{u}_i)u_i^2\}\nonumber\\
&&{}+\sum_{j\neq i}^N u_{i\alpha}(v_{\beta}(\mathbf{r},t)v_{\gamma}
(\mathbf{r}',t)+v_{\beta}(\mathbf{r},t)u_{j\gamma}+u_{i\beta}v_{\gamma}
(\mathbf{r}',t)+u_{i\beta}u_{j\gamma})G^{\beta\gamma}_{ij}\biggr].
\end{eqnarray}

Thermal work density $\alpha(\mathbf{r},t)$ (which is part of work
density producing by particles in thermal motion) in the energy
balance equation (\ref{fla16}) is defined in the
next way:
\begin{eqnarray}
\label{fla12}
&&\alpha(\mathbf{r},t)=\frac{1}{\Delta}\int_{\Delta(\mathbf{r})}d\xi\sum_{i,j=1,i\neq j}^N\delta(\mathbf{r}+\xi-\mathbf{r}_i(t))\biggl[-\frac{e^2}{2}(u_{i\alpha}+u_{j\alpha})
\partial_{i\alpha}G_{ij}\nonumber\\
&&{}+\frac{e^2}{4c^2}\partial_{i\gamma}G^{\alpha\beta}_{ij}[(v_{\gamma}(\mathbf{r},t)+
v_{\gamma}(\mathbf{r}',t)+u_{i\gamma}+u_{j\gamma})(v_{\alpha}(\mathbf{r},t)+
u_{i\alpha})\times\nonumber\\
&&{}\times(v_{\beta}(\mathbf{r}',t)+u_{j\beta})
-(v_{\gamma}(\mathbf{r},t)+v_{\gamma}(\mathbf{r}',t))v_{\alpha}(\mathbf{r},t)
v_{\beta}(\mathbf{r}',t)]\nonumber\\
&&{}+\frac{e^3}{4mc^2}G^{\alpha\beta}_{ij}(E_{i\alpha}u_{j\beta}-E_{j\beta}
u_{i\alpha})\biggr].
\end{eqnarray}

The thermal energy current $q^\alpha(\mathbf{r},t)$ and the
thermal work density $\alpha(\mathbf{r},t)$, unlike the
non-relativistic case, include the velocity field
$v^\alpha(\mathbf{r},t)$.

Thus, equations of the five-moment approximation in
semi-relativistic hydrodynamics were obtained. With method
described above this equations were obtained for the first time.

\section{Quantum hydrodynamics. Formulation of problem in semi-relativistic approximation}

Let us consider a quantum mechanical system of N charged
particles, interacting by the Coulomb and the current-current
interactions, placed into an external electromagnetic field. Derivation of quantum hydrodynamical
equations is carried out by the method described in Refs.
\cite{LSK1999}, \cite{LSK2001}, \cite{Andreev LSK Trukhanova PRB}.
Microscopic mass density is defined by formula
\begin{equation}
\label{fla2}
\rho(\mathbf{r},t)=\int{dR}\sum_{i=1}^N m_i \delta(\mathbf{r}-\mathbf{r}_i)\psi^*(R,t)\psi(R,t),
\end{equation}
where $R=(\mathbf{r}_1,...,\mathbf{r}_N)$, $\mathbf{r}_i$ - coordinates of $i$-th particle,
\begin{equation}dR=\prod_{j=1}^N d\mathbf{r}_j,\end{equation}
$dR$ is an element of volume in 3N dimensional
configuration space, $d\mathbf{r}_j$ is the element of
volume in three dimensional space of radius-vector $\mathbf{r}_j$.
Number density and charge density are defined in the
same way. Formula (\ref{fla2}) appears as the quantum mechanical averaging \cite{Landau Vol 3} of the mass density operator giving microscopic quantum observable suitable for the collective effect description.

The Hamiltonian of system has form ($\hat H_0$ is non-relativistic part of the Hamiltonian, $\hat H_r$ is semi-relativistic part):
\begin{equation}\label{Hamiltonian full as sum} \hat H=\hat H_0+\hat H_r,\end{equation}
\begin{equation}\label{Hamiltonian non rel part} \hat H_0=\sum_{i=1}^N\left(\frac{\mathbf{D}_i ^2}{2m_i}+e_i \varphi_i\right)+\frac{1}{2}\sum_{i,j=1,i\neq j}^{N}e_i e_j \ G_{ij},\end{equation}

\begin{equation}
\label{fla4}
\hat H_r=-\sum_{i=1}^N\frac{\mathbf{D}_i^4}{8m_i^3c^2}-\sum_{i,j=1,i\neq j}^{N} \frac{e_i e_j}{4m_i m_j c^2}\ G_{ij}^{\alpha\beta}D_{i}^{\alpha}D_{j}^{\beta},
\end{equation}
where $D_{i}^\alpha=(\hbar/i)\ \partial_{i}^\alpha-(e_i/c)A_{i}^\alpha$, $\partial_{i}^\alpha$ is derivative over coordinates of $i$-th particle, functions $\varphi_i=\varphi_i(\mathbf{r}_i,t)$, $A_i^\alpha=A_i^\alpha(\mathbf{r}_i,t)$ are the potentials of external electromagnetic field, $\mathbf{r}_{ij}=\mathbf{r}_i-\mathbf{r}_j$. $G_{ij},~G_{ij}^{\alpha\beta}$ are the Green functions of the Coulomb and the current-current interactions respectively, defining in the next way:
\begin{equation}G_{ij}=\frac{1}{r_{ij}},~ G_{ij}^{\alpha\beta}=\frac{\delta^{\alpha\beta}}{r_{ij}}+
\frac{x^{\alpha}_{ij}x^{\beta}_{ij}}{r_{ij}^3}.\end{equation}

Previously, in section \ref{classic_hd}, with method of differentiation the equations (\ref{cont eq classicic})-(\ref{fla12}) were derived for classical case. In sections \ref{continuity_eq}-\ref{intro_vel_field} equations of quantum hydrodynamics is derived.

\section{Continuity equation}\label{continuity_eq}

Differentiating mass density (\ref{fla2})  with
respect to time and using the Schr\"{o}dinger equation
with the Hamiltonian (\ref{Hamiltonian full as sum}), the continuity equation is
obtained:
\begin{equation}\partial_t\rho+\partial_\alpha j^{\alpha}=0,\end{equation}
where the quantum particle current $j^\alpha$ has form
\begin{eqnarray}
\label{fla3}
j^{\alpha}(\mathbf{r},t)&=&\int{dR}\sum_{i=1}^N \delta(\mathbf{r}-\mathbf{r}_i)\biggl[\frac{1}{2}\psi^* D_{i}^{\alpha}\psi\nonumber\\
&&{}-\frac{1}{8m_i^2c^2}(\psi^*D_{i}^{\alpha}D_i^2\psi+D_{i}^{\alpha}\psi
D_i^{*2}\psi^*)\nonumber\\
&&{}-\sum_{j=1,j\neq i}^N \frac{e_i e_j}{4m_j c^2}G_{ij}^{ \alpha\beta}\psi^*D_{j}^{\beta}\psi+c.~c.\biggr].
\end{eqnarray}

Form of the continuity equation in comparison with the
non-relativistic case does not change, but expression for the
current obtains corresponding corrections.

\section{Euler equation}

In the same way, differentiating with respect to time expression
for current (\ref{fla3}), we obtain the quantum equation of
particle current evolution (the Euler equation):
\begin{equation}\label{Euler quantum general form}\partial_t j^\alpha+\partial_{\beta}\Pi^{\alpha\beta}=en E^\alpha_{ext}+\frac{e}{mc}\ \varepsilon^{\alpha\beta\gamma}j^\beta
B^\gamma_{ext}+\mathcal{F}^\alpha.\end{equation}

General form of this equation coincides with classical analog
(\ref{classic momentum bal eq}). However, functions $\Pi^{\alpha\beta}$
and $\mathcal{F}_\alpha$ contain quantum contribution. The quantum flux of
particle current $\Pi^{\alpha\beta}(\mathbf{r},t)$ equals to
\begin{eqnarray}
\label{tensor_mom_cur}
&&\Pi^{\alpha\beta}=\int{dR}\sum_{i=1}^N \delta(\mathbf{r}-\mathbf{r}_i)\biggl[\frac{1}{4m_i}(\psi^*D_{i}^{\beta}D_i^\alpha\psi+
D_i^{*\beta}\psi^*D_i^\alpha\psi)\nonumber\\
&&{}-\frac{1}{8m_i^3c^2}(\psi^*D_i^\beta D_i^\alpha D_i^2\psi+D_i^{*\beta}\psi^*D_i^\alpha D_i^2\psi+D_i^{*\alpha}\psi^*D_i^\beta D_i^2\psi\nonumber\\
&&{}+D_i^{*2}\psi^*D_i^\beta D_i^\alpha\psi)-
\sum_{j=1,j\neq i}^N\frac{e_i e_j}{8m_i m_j c^2}[G_{ij}^{\beta\gamma}(\psi^*D_j^\gamma D_i^\alpha\psi+D_j^{*\gamma}\psi^*D_i^\alpha\psi)\nonumber\\
&&{}+G^{\alpha\gamma}_{ij}
(\psi^*D_j^\gamma D_i^\beta\psi+D_j^{*\gamma}\psi^*D_i^\beta\psi)]+c.~c.\biggr].
\end{eqnarray}
In non-relativistic limit the flux of
particle current coincides with the momentum current, but in semi-relativistic and full relativistic cases this is no longer true \cite{Andreev arxiv rel}.

As in the case of the current, the tensor of flux of the
particle current
contains corrections appearing in the semi-relativistic
approximation. It is possible to show that this tensor is
symmetrical:
$\Pi^{\alpha\beta}(\mathbf{r},t)=\Pi^{\beta\alpha}(\mathbf{r},t)$.

Force density $\mathcal{F}_\alpha(\mathbf{r},t)$ can be
represented as sum of three terms, which reflects contribution
from various types of interaction:
\begin{equation}
\label{fla17}
\mathcal{F}^\alpha(\mathbf{r},t)=\mathcal{F}^\alpha_{cl}(\mathbf{r},t)+
\mathcal{F}^\alpha_r(\mathbf{r},t)+\mathcal{F}^\alpha_{cur}(\mathbf{r},t),
\end{equation}
where
\begin{equation} \label{cl force quantum}
\mathcal{F}^\alpha_{cl}=-\int dR\sum_{i,j=1,i\neq j}^N\delta(\mathbf{r}-\mathbf{r}_i)e_i e_j \partial_{i}^{\alpha}G_{ij}\psi^*\psi,
\end{equation}
describes contribution of the Coulomb interaction in the force density,
\begin{eqnarray} \label{rel correct to kin en in force field}
&&\mathcal{F}^\alpha_r=-\int dR\sum_{i=1}^N\delta(\mathbf{r}-\mathbf{r}_i)\frac{e_i}{4m_i^2c^2}[E_{i}^\alpha
(\psi^*D_i^2\psi+c.~c.)\nonumber\\
&&{}+E_{i}^\beta(\psi^*D_{i}^{\beta}D_i^\alpha\psi+D_i^{*\beta}\psi^*D_i^\alpha\psi+c.~c.)-
\hbar^2\partial_i^{\beta}(\partial_i^{\alpha}E_{i}^\beta\psi^*\psi)]
\end{eqnarray}
is part of force density associated with the relativistic correction to kinetic energy. This quantity consists from three groups of terms. First of them is proportional to kinetic energy density $\mathcal{F}^\alpha_{r,\varepsilon}=-\frac{e}{mc^{2}}E^{\alpha}\varepsilon$, second is proportional to the flux of particle current $\mathcal{F}^\alpha_{r,\Pi}=-\frac{e}{mc^{2}}E^{\beta}\Pi^{\alpha\beta}$, third gives new purely quantum term in the Euler equation. It is proportional to a divergence of tensor, which consists from product of number density on spatial derivative of electric field $\mathcal{F}^\alpha_{r,\hbar}=\frac{e\hbar^{2}}{4m^{2}c^{2}}\partial^{\beta}(n\partial^{\alpha}E^{\beta})$.

Third term in equation (\ref{fla17}) has form
\begin{eqnarray} \label{F curr quant}
&&\mathcal{F}^\alpha_{cur}=\int{dR}\sum_{i,j=1,i\neq j}^N\delta(\mathbf{r}-\mathbf{r}_i)\biggl[-\frac{e_i e_j^2}{2m_j c^2}G^{\alpha\beta}_{ij}E^{\beta}_{j}\psi^*\psi\nonumber\\
&&{}+\frac{e_i e_j}{8m_i m_j c^2}(\partial_{i}^\alpha G^{\beta\gamma}_{ij}-
\partial_{i}^\beta G^{\alpha\gamma}_{ij})(\psi^*D_{i}^{\beta}D_j^\gamma\psi+
D_i^{*\beta}\psi^*D_j^\gamma\psi+c.~c.)\nonumber\\
&&{}+\frac{e_i e_j}{8m_j^2c^2}\partial_{i}^\gamma G^{\alpha\beta}_{ij}
(\psi^*D_{j}^{\beta}D_j^\gamma\psi+D_j^{*\beta}\psi^*D_j^\gamma\psi+c.~c.)\nonumber\\
&&{}-\frac{e_i e_j\hbar^2}{8m_i m_j c^2}\partial_{i}^\alpha G^{\beta\gamma}_{ij}\partial_i^{\beta}\partial_j^{\gamma}(\psi^*\psi)\biggr],
\end{eqnarray}
and represents part of field density associated with the
current-current interaction between particles. It consists from
four groups of terms. The first group corresponds to existing in
classical theory term arising from explicit dependence of particle
momentum from velocities of other particles ($\mathbf{p}_i\neq
m_i\mathbf{v}_i$). The second group being considered in the self-consistent field approximation, with dropping thermal contribution as well, leads to magnetic part of the Lorentz force $e[\mathbf{j},\mathbf{B}^{int}]/(mc)$ caused by the current-current interaction (we show it explicitly in the Appendix). The third group is proportional to
the flux of particle current and corresponds to the classical expression
(see the third line in formula (\ref{fla5}). The fourth group of terms is totally
quantum, it has complex structure and, in particular, is
proportional to second derivative of two-particle concentration
(\ref{fla18}).

Full electric field acting on $i$-th particle is defined in next
way:
\begin{equation}E_{i\alpha}=E_{i,ext}^\alpha-\sum_{j=1,j\neq i}^N e_j \partial_i^{\alpha}G_{ij}.\end{equation}

The Euler equation, unlike the continuity equation, interaction
enters in explicit form. In case of system with the Coulomb
interaction obtained equation coincides by form with classical
equation, however, it contains quantum correlations caused by
exchange interactions. Quantum character shows in the presence of
additional quantum pressure called the quantum Bohm potential.

At account of the semi-relativistic corrections, essentially new
terms appear in the Euler equation. Two group of terms appear in
all: first associated with the relativistic correction to kinetic
energy, second - with the current-current interaction. Part of
these terms is classical, i. e. appears at constructing of
classical hydrodynamics based on the Darwin Lagrangian (but we
should note that the flux of particle current and the energy density
contain quantum parts, which is absent in the classical theory).
Essentially new terms are the quantum terms explicitly
proportional to $\hbar$.

Deriving equations of quantum hydrodynamics we start from
definition of number (or mass) density, its differentiation gives
continuity equation. In this equation expression for current
density through wave function appears. In equation for current
density new functions appear, for which by differentiation over
time evolution equations can be derived. In this way infinite
chain of equations in 3D physical space is obtained, similar to
BBGKY chain in kinetics. This chain of equations is equivalent to
the initial Schr\"{o}dinger equation in configuration space.

Equations obtained above are written in general form. We consider these equations in self-consistent field approximation further. If we introduce two-particle concentration in the Coulomb force field (\ref{cl force quantum})
\begin{equation}\label{fla18}n_2(\mathbf{r},\mathbf{r}',t)=\int{dR}\sum_{i,j=1,i\neq j}^N \delta(\mathbf{r}-\mathbf{r}_i)\delta(\mathbf{r}'-\mathbf{r}_j)\psi^*(R,t)\psi(R,t),
\end{equation}
then in case of system consisting from one kind of particles, and also in self-consistent field approximation
($n_2(\mathbf{r},\mathbf{r}',t)=n(\mathbf{r},t)n(\mathbf{r}',t)$), expression for force density of the Coulomb interaction $\mathcal{F}^\alpha_{cl}$ take form
\begin{equation}\mathcal{F}^\alpha_{cl}(\mathbf{r},t)=-e^2n(\mathbf{r},t)
\ \partial^{\alpha}\int G(\mathbf{r}-\mathbf{r}')n(\mathbf{r}',t)d\mathbf{r}'.\end{equation}

Further, after introduction of velocity field in section \ref{intro_vel_field}, all quantities in obtained equations are written in the self-consistent field approximation.

\section{Energy balance equation}

Investigating hydrodynamical equations five-moment
approximation is considered usually, consisting of continuity
equation, momentum balance equation and energy balance equation.
Continuity equation and momentum balance equation were obtained
above. Now we proceed to derivation of quantum energy balance
equation. Energy density for considered system of particles is
defined in next way:
\begin{eqnarray}
&&\varepsilon(\mathbf{r},t)=\int dR\sum_{i=1}^N\delta(\mathbf{r}-\mathbf{r}_i)\biggl[\biggl(\frac{1}{4m_i}\psi^*D_i^2\psi-\frac{1}
{16m_i^3c^2}\psi^*D_i^4\psi+c.~c.\biggr)\nonumber\\
&&{}+\sum_{j=1,j\neq i}^N\biggl(\frac{1}{2}e_i e_j G_{ij}\psi^*\psi-\frac{e_i e_j}{8m_i m_j c^2}G^{\alpha\beta}_{ij}(\psi^*\dia D_j^\beta\psi+c.~c.)\biggr)\biggr].
\end{eqnarray}

As in the case of mass current density (\ref{fla3}), differentiating this expression over time and using Schr\"{o}dinger equation we obtain energy balance equation:
\begin{equation}\label{energy_eq}
\partial_t \varepsilon(\mathbf{r},t)+\partial_\alpha Q^\alpha(\mathbf{r},t)=j_e^\alpha(\mathbf{r},t) E_\alpha(\mathbf{r},t)+\mathcal{A}(\mathbf{r},t),
\end{equation}
where $Q^\alpha(\mathbf{r},t)$ is vector of energy current density, $\mathcal{A}(\mathbf{r},t)$ is work density. Vector of energy current density has complex structure. We divide it on three parts
\begin{equation}
\label{fla19}
Q^\alpha(\mathbf{r},t)=Q^\alpha_0(\mathbf{r},t)+Q^\alpha_{coul}(\mathbf{r},t)+
Q^\alpha_{r,cur}(\mathbf{r},t),
\end{equation}
and consider them separately. First term in (\ref{fla19}) is non-relativistic kinetic energy current density (see \cite{LSK1999}) and has form
\begin{equation}
Q^\alpha_0=\int dR\sum_{i=1}^N\delta(\mathbf{r}-\mathbf{r}_i)\frac{1}{8m_i^2}(\psi^*D_i^\alpha D_i^2\psi+D_i^{*\alpha}\psi^*D_i^2\psi+c.~c.).
\end{equation}
Energy current of the Coulomb interaction is represented by second term in formula (\ref{fla19}):
\begin{eqnarray}
&&Q^\alpha_{coul}=\int dR\sum_{i,j=1,i\neq j}^N\delta(\mathbf{r}-\mathbf{r}_i)e_i e_j G_{ij}\biggl[\frac{1}{4m_i}(\psi^*D_i^{\alpha}\psi+c.~c.)\nonumber\\
&&{}-\frac{1}{16m_i^3c^2}(\psi^*D_i^\alpha D_i^2\psi+D_i^{*\alpha}\psi^*D_i^2\psi+c.~c.)\nonumber\\
&&{}-\sum_{k=1,k\neq j}^N\frac{e_i e_k}{8m_i m_k c^2}G_{ik}^{\alpha\beta}(\psi^*D_k^{\beta}\psi+c.~c.)\biggr],
\end{eqnarray}
where expression in square brackets is analogous by its structure to the current $j^\alpha$ (\ref{fla3}) and consists from three parts. First part corresponds to result obtained earlier from non-relativistic theory \cite{LSK1999}, second and third parts are caused by contribution from semi-relativistic effects (the addition to kinetic energy and the current-current interaction, correspondingly).

Part of energy current $Q^\alpha_{r,cur}$ arising from semi-relativistic corrections in the Hamiltonian (\ref{fla4}) is defined by expression
\begin{eqnarray}
&&Q^\alpha_{r,cur}=-\int dR\sum_{i=1}^N\delta(\mathbf{r}-\mathbf{r}_i)\biggl[\frac{1}{16m_i^4c^2}(\psi^*D_i^\alpha D_i^4\psi\nonumber\\
&&{}+D_i^{*\alpha}\psi^*D_i^4\psi+D_i^{*2}\psi^*D_i^\alpha D_i^2\psi+c.~c.)\nonumber\\
&&{}+\sum_{j=1,j\neq i}^N\frac{e_i e_j}{16m_i^2 m_j c^2}
\{G^{\alpha\beta}_{ij}(\psi^*D^\beta_j D_i^2\psi+D_j^{*\beta}\psi^*D_i^2\psi+c.~c.)\nonumber\\
&&{}+G^{\beta\gamma}_{ij}(\psi^*D_i^\alpha D_i^\beta D_j^\gamma\psi+D_i^{*\alpha}\psi^*D_i^\beta D_j^\gamma\psi+c.~c.)\}\biggr].
\end{eqnarray}
Terms in the first two lines arise from the semi-relativistic correction to kinetic energy, in the last two lines - from the current-current interaction between particles.

Expression for work density $\mathcal{A}(\mathbf{r},t)$ can be also divided in three parts:
\begin{equation}\mathcal{A}(\mathbf{r},t)=\mathcal{A}^{cl}(\mathbf{r},t)+
\mathcal{A}^r(\mathbf{r},t)+
\mathcal{A}^{cur}(\mathbf{r},t),\end{equation}
where
\begin{eqnarray}&&\mathcal{A}^{cl}=\int dR\sum_{i,j=1,j\neq i}^N\delta(\mathbf{r}-\mathbf{r}_i)e_i e_j\partial_i^\alpha G_{ij}\biggl[
\frac{1}{4m_i}(\psi^*D_i^\alpha \psi+c.~c.)\nonumber\\
&&{}-\frac{1}{4m_j}(\psi^*D_j^\alpha \psi+c.~c.)
-\frac{1}{16m_i^3c^2}(\psi^*D_i^\alpha D_i^2\psi+D_i^{*\alpha}\psi^*D_i^2\psi\nonumber\\
&&{}+c.~c.)
+\frac{1}{16m_j^3c^2}(\psi^*D_j^\alpha D_j^2\psi+D_j^{*\alpha}\psi^*D_j^2\psi+c.~c.)\nonumber\\
&&{}-\sum_{k=1,k\neq i}^N\frac{e_i e_k}{8m_i m_k c^2}G_{ik}^{\alpha\beta}(\psi^*D_k^\beta \psi+c.~c.)\nonumber\\
&&{}+\sum_{k=1,k\neq j}^N\frac{e_j e_k}{8m_j m_k c^2}G_{jk}^{\alpha\beta}(\psi^*D_k^\beta \psi+c.~c.)\biggr],
\end{eqnarray}
$\mathcal{A}^{cl}$ is the Coulomb part of work density (it is seen from the definition of mass current density (\ref{fla3}) and formula
$j^\alpha(\mathbf{r},t)=\int dR\sum_{i=1}^N\delta(\mathbf{r}-\mathbf{r}_i)\psi^*\psi m_i v_{i\alpha}$,
from which it is seen that
\begin{eqnarray}
\via&=&\frac{1}{2m_i}\frac{\dia\psi}{\psi}-\frac{1}{8m_i^3c^2}\frac{1}{\psi^*\psi}
(\psi^*\dia\dis\psi+\diac\psi^*\dis\psi)\nonumber\\
&&{}-\sum^N_{j=1,j\neq i}\frac{e_i e_j}{4m_i m_j c^2}G_{ij}^{\alpha\beta}\frac{\djb\psi}{\psi}+c.~c.,
\end{eqnarray}
after this it is easy to see correspondence between formula for $\mathcal{A}^{cl}$ and corresponding part of classical energy balance equation (\ref{fla16})). Work density $\mathcal{A}^{r}$ associated with the relativistic correction to kinetic energy looks in the following way:
\begin{eqnarray}&&\mathcal{A}^{r}=\int dR\sum_{i=1}^N\delta(\mathbf{r}-\mathbf{r}_i)\frac{ e_i}{8m_i^3c^2}\biggl[\hbar^2\partial_i^\alpha[\partial_i^\beta E_i^\beta(\psi^*D_i^{\alpha}\psi+c.~c.)]\nonumber\\
&&{}+\hbar^2\Delta_i E_i^\alpha(\psi^*D_i^{\alpha}\psi+c.~c.)+\hbar^2 E_i^\alpha\partial_i^\alpha\partial_i^\beta (\psi^*D_i^{\beta}\psi+c.~c.)\nonumber\\
&&{}-2\frac{\hbar}{i}\partial_i^\alpha E_i^\beta(\psi^*D_i^{\alpha}D_i^{\beta}\psi-c.~c.)\biggr].
\end{eqnarray}
This function is proportional to square of Planck constant $\hbar$ (with respect to fact that expression in the last line also gives square of $\hbar$ at expansion). Terms in the first two lines are proportional to derivative of electric field and current density, and with respect to this fact this expression (in the case of system with one kind of particles) can be written in the following form:
\begin{eqnarray}
&&\mathcal{A}^r=\frac{e\hbar^2}{4m^3c^2}\biggl[\partial_\alpha(\partial_\beta E_\beta j^\alpha)+
\Delta E_\alpha j^\alpha+E_\alpha\partial_\alpha(\partial_\beta j^\beta)\nonumber\\
&&{}+\partial_\alpha E_\beta\int dR\sum_{i=1}^N\delta(\mathbf{r}-\mathbf{r}_i)\frac{i}{\hbar}
(\psi^*D_i^{\alpha}D_i^{\beta}\psi-c.~c.)\biggr].
\end{eqnarray}

Finally we obtain work density $\mathcal{A}^{cur}$ associated with the current-current interaction between particles:
\begin{eqnarray}&&\mathcal{A}^{cur}=\int dR\sum_{i,j=1,j\neq i}^N\delta(\mathbf{r}-\mathbf{r}_i)\frac{e_i e_j}{8m_i m_j c^2}[e_i E_i^\alpha G_{ij}^{\alpha\beta}(\psi^*D_j^\beta \psi+c.~c.)\nonumber\\
&&{}-e_j E_j^\beta G_{ij}^{\alpha\beta}(\psi^*D_i^\alpha \psi+c.~c.)+\frac{2}{m_i}\partial_i^\gamma G^{\alpha\beta}_{ij}(\psi^*D_i^\gamma D_i^\alpha D_j^\beta\psi+c.~c.)\nonumber\\
&&{}-\frac{1}{2m_i}\partial_i^\gamma G^{\alpha\beta}_{ij}
(\psi^*D_i^\gamma D_i^\alpha D_j^\beta\psi+D_i^{*\gamma}\psi^*D_i^\alpha D_j^\beta\psi+c.~c.)\nonumber\\
&&{}+\frac{1}{2m_j}\partial_i^\gamma G^{\alpha\beta}_{ij}
(\psi^*D_j^\gamma D_i^\alpha D_j^\beta\psi+D_j^{*\gamma}\psi^*D_i^\alpha D_j^\beta\psi+c.~c.)\nonumber\\
&&{}+\frac{\hbar}{i m_i}\Delta_i G^{\alpha\beta}_{ij}(\psi^*D_i^\alpha D_j^\beta\psi-c.~c.)].
\end{eqnarray}
Expression in the last line is completely quantum. Other terms correspond to the classical current-current terms in the right-hand side of energy balance equation (\ref{fla16}), but as in the case of tensor of the flux of particle current (\ref{tensor_mom_cur}) they give a quantum contribution. So we have an analog of the quantum Bohm potential.

\section{Introduction of velocity field}\label{intro_vel_field}

\subsection{Exponential form of wave function and velocity of particle}

Basic fact is the fact that mass current density is $j_\alpha(\mathbf{r},t)=mn(\mathbf{r},t)v_\alpha(\mathbf{r},t)$, i. e. velocity field $v_\alpha(\mathbf{r},t)$ is defined from obtained above expressions for number density and current density.

Wave function of N particle system can be represented as
\begin{equation}\label{fla6}\psi(R,t)=a(R,t)\exp(iS(R,t)/\hbar),\end{equation}
In formula (\ref{fla6}), which is called the Madelung decomposition, $a^2(R,t)$ is the probability density of detection in volume $(R,R+dR)$ of 3N dimensional configuration space, and phase of wave function $S(R,t)$ is quantity, gradient of which corresponds to probability current in non-relativistic case.

Substituting given expression (\ref{fla6}) in the formula for the current (\ref{fla3}) we obtain:
\begin{equation}j^\alpha(\mathbf{r},t)=\int dR\sum_{i=1}^N\delta(\mathbf{r}-\mathbf{r}_i)m_i a^2v_i^\alpha,\end{equation}
where
\begin{eqnarray}\label{velocity quantum}
v_i^\alpha&=&\frac{s_i^\alpha}{m_i}-\frac{s_i^\alpha s_i^2}{2m_i^3 c^2}
+\frac{\hbar^2}{2m_i^3c^2}\biggl[s_i^\alpha a^{-1}\Delta_i a+\partial_i^\alpha(s_i^\beta\partial_i^\beta \ln a)\nonumber\\
&&{}+\frac{1}{2}\partial_i^\alpha\partial_i^\beta s_i^\beta\biggr]
-\sum_{j=1,j\neq i}^N\frac{e_i e_j}{2m_i m_j c^2}G^{\alpha\beta}_{ij}s_j^\beta,
\end{eqnarray}
and quantity $\textbf{s}_i$ is defined as $s_i^\alpha=\partial_i^\alpha S-\frac{e_i}{c}A_i^\alpha$. Velocity of $i$-th particle should be divided into the velocity field and the thermal velocity of of $i$th particle: $v_i^\alpha(R,t)=v^\alpha(\mathbf{r},t)+u_i^\alpha(\mathbf{r},R,t)$. The velocity field is introduced in order that the current equals to zero on the thermal velocities:
\begin{equation}\int dR\sum_{i=1}^N\delta(\mathbf{r}-\mathbf{r}_i)m_i a^2u_i^\alpha=0.\end{equation}
As a result we obtain familiar expression for the mass current density (for convenience we consider system with one kind of particles, i. e. $m_i=m,~ e_i=e$ for each $i$):
\begin{equation}\label{velocity and j relation} j^\alpha(\mathbf{r},t)=mn(\mathbf{r},t)v^\alpha(\mathbf{r},t).\end{equation}
In terms of the velocity field, the continuity equation takes the well-known form form
\begin{equation}
\label{fla13}
\partial_t n(\mathbf{r},t)+\partial_\alpha[n(\mathbf{r},t)v^\alpha(\mathbf{r},t)]=0.
\end{equation}

\subsection{Equation for the velocity field evolution}

Using the Madelung decomposition for wave function (\ref{fla6}) current and energy evolution equations can be expressed through familiar hydrodynamical and thermodynamical quantities.

For system with one kind of particles the current evolution equation takes form
\begin{eqnarray}
\label{fla14}
&&mn(\mathbf{r},t)(\partial_t +v_\beta(\mathbf{r},t)\partial_\beta )v_\alpha(\mathbf{r},t)+\partial_\beta p_{\alpha\beta}(\mathbf{r},t)+\partial_\beta T_{\alpha\beta}(\mathbf{r},t)=\nonumber\\
&&{}=enE_\alpha+\frac{e}{c}n\varepsilon_{\alpha\beta\gamma}v_\beta B_\gamma+
\mathcal{F}_\alpha^{cl}+\mathcal{F}_\alpha^{q},
\end{eqnarray}
where $p_{\alpha\beta}(\mathbf{r},t)$ is the kinetic pressure tensor, $T_{\alpha\beta}(\mathbf{r},t)$ is the quantum addition to the thermal pressure (the quantum Bohm potential), the electric field $\textbf{E}$ is the sum of the external field and the field associated with the interaction between particles: $E_\alpha(\mathbf{r},t)=E_\alpha^{ext}(\mathbf{r},t)+E_\alpha^{int}(\mathbf{r},t)$, and similarly for the magnetic field:
$B_\alpha(\mathbf{r},t)=B_\alpha^{ext}(\mathbf{r},t)+B_\alpha^{int}(\mathbf{r},t).$
Function $\mathcal{F}_\alpha^{cl}$ is the classical semi-relativistic part of force density (see (\ref{fla5}); one needs to account the fact that functions $p_{\alpha\beta}(\mathbf{r},t)$ and $\rho\epsilon(\mathbf{r},t)$ contain quantum part, for example, instead of $p_{\alpha\beta}$ it should be written $p_{\alpha\beta}+T_{\alpha\beta}$), and $\mathcal{F}_\alpha^{q}$ is a specific part of the force density appearing in the QHD:
\begin{eqnarray} \label{Force q with vel f}
&&\mathcal{F}_\alpha^{q}=\frac{e\hbar^2}{4m^2c^2}\partial_\beta(\partial_\alpha E_\beta\cdot n)\nonumber\\
&&{}-\frac{e^2\hbar^2}{8m^2c^2}\partial_\beta n(\mathbf{r},t)\int d\mathbf{r}'\partial_\alpha G_{\beta\gamma}(\mathbf{r}-\mathbf{r}')\partial^\prime_\gamma
n(\mathbf{r}',t).
\end{eqnarray}

Fields $\mathbf{E}^{int}$, $\mathbf{B}^{int}$ associated with interaction between particles of system satisfy to following formulas:
\begin{equation}\mathbf{E}^{int}(\mathbf{r},t)=-\mathbf{\nabla}\varphi^{int}(\mathbf{r},t),
\end{equation}
and
\begin{equation}\mathbf{B}^{int}(\mathbf{r},t)=\mathrm{curl}\mathbf{A}^{int}(\mathbf{r},t),
\end{equation}
with
\begin{equation}\varphi^{int}(\mathbf{r},t)=e\int d\mathbf{r}'G(\mathbf{r}-\mathbf{r}')n(\mathbf{r}',t),\end{equation}
and
\begin{equation}A^{int}_\alpha(\mathbf{r},t)=\frac{e}{2c}\int d\mathbf{r}'G_{\alpha\beta}(\mathbf{r}-\mathbf{r}')n(\mathbf{r}',t)v_\beta(\mathbf{r}',t).
\end{equation}
These fields obey the quasi-static Maxwell equations:
\begin{equation}\label{QRD divE}\nabla \textbf{E}(\textbf{r},t)=4\pi \sum_{a}e_{a}n_{a}(\textbf{r},t),
\end{equation}
\begin{equation}\label{QRD curlE}\nabla\times \textbf{E}(\textbf{r},t)=0,
\end{equation}
\begin{equation}\label{QRD divB}\nabla \textbf{B}(\textbf{r},t)=0,
\end{equation}
and
\begin{equation}\label{QRD curlB}\nabla\times \textbf{B}(\textbf{r},t)=\frac{4\pi}{c}\sum_{a}e_{a}n_{a}(\textbf{r},t)\textbf{v}_{a}(\textbf{r},t),
\end{equation}
where $a$ stands for species of particles.

Tensor $\pi_{\alpha\beta}(\mathbf{r},\mathbf{r}',t)$ entering in $\mathcal{F}_\alpha^{cl}$ (\ref{fla5}) looks like
\begin{eqnarray}\label{pialphabeta}&&\pi_{\alpha\beta}(\mathbf{r},\mathbf{r}',t)=\\
&&{}~~~=\int dR\sum_{i,j=1,i\neq j}^N\delta(\mathbf{r}-\mathbf{r}_i)\delta(\mathbf{r}'-\mathbf{r}_j)a^2\biggl(u_{i\alpha}
u_{j\beta}-\frac{\hbar^2}{2m^2}\partial_{i\alpha}\partial_{j\beta}\ln a\biggr).\nonumber
\end{eqnarray}

Tensor $T_{\alpha\beta}(\mathbf{r},t)$ corresponding to the quantum Bohm potential has form
\begin{eqnarray}
\label{Bohm potential}
&&T_{\alpha\beta}(\mathbf{r},t)=\int dR\sum_{i=1}^N\delta(\mathbf{r}-\mathbf{r}_i)a^2\biggl[-\frac{\hbar^2}{2m}\biggl
(1-\frac{v_i^2}{c^2}\biggr)\partial_{i\alpha}\partial_{i\beta}\ln a\nonumber\\
&&{}+\frac{\hbar^2}{2m c^2}(\partial_{i\alpha}v_{i\gamma}\partial_{i\beta}v_{i\gamma}+v_{i\gamma}
\partial_{i\alpha}\partial_{i\beta}v_{i\gamma})\nonumber\\
&&{}+\frac{\hbar^2}{4m c^2}(v_{i\alpha}\partial_{i\beta}+v_{i\beta}\partial_{i\alpha})(\partial_{i\gamma}v_{i\gamma}+
2v_{i\gamma}\partial_{i\gamma}\ln a)\nonumber\\
&&{}-\frac{\hbar^4 a^{-2}}{4m^3 c^2}(a\partial_{i\alpha}\partial_{i\beta}\Delta_i a+\partial_{i\alpha}\partial_{i\beta}a\Delta_i a-\partial_{i\alpha}a\partial_{i\beta}\Delta_i a-\partial_{i\beta}a\partial_{i\alpha}\Delta_i a)\biggr]\nonumber\\
&&{}+\int dR\sum_{i=1,j=1,i\neq j}^N\delta(\mathbf{r}-\mathbf{r}_i)a^2\frac{\hbar^2e^2}{4m^2c^2}(G^{\beta\gamma}_{ij}
\partial_{i\alpha}\partial_{j\gamma}\ln a+G^{\alpha\gamma}_{ij}
\partial_{i\beta}\partial_{j\gamma}\ln a).
\end{eqnarray}
This is a general representation of the semi-relativistic quantum Bohm potential. It is defined in terms of the amplitude $a=a(R,t)$ of many-particle wave function (\ref{fla6}) and the quantum velocities $\textbf{v}_{i}$ appearing via the phase of many-particle wave function (\ref{velocity quantum}). Usually, at applications of quantum hydrodynamics, the quantum Bohm potential is used in approximation of independent particles, see for instance Ref. \cite{Andreev LSK Trukhanova PRB} formulas (10)-(11), and Ref. \cite{Shukla Eliasson 2011} formula (17). Below we present approximate form of the semi-relativistic quantum Bohm potential (\ref{Bohm potential}). The first term in the first group of terms in (\ref{Bohm potential}) is the non-relativistic part of the Bohm potential obtained in Ref. \cite{LSK1999} (see formula (35)). Its approximate form is
\begin{equation}\label{SRW Bohm}T^{\alpha\beta}_{nr}=-\frac{\hbar^{2}}{4m}\partial^{\alpha}\partial^{\beta}
n+\frac{\hbar^{2}}{4m}\Biggl(\frac{\partial^{\alpha}n\cdot\partial^{\beta}n}{n}\Biggr).\end{equation}
We should also represent it as the density of force field
\begin{equation}\label{SRW Bohm}F_{Q}^{\alpha}=-\partial_{\beta}T^{\alpha\beta}=\frac{\hbar^{2}}{2m}n
\partial_{\alpha}\frac{\triangle\sqrt{n}}{\sqrt{n}}\end{equation}
to make it more recognizable.

The full semi-relativistic Bohm potential arises in approximation of independent particles as
$$T^{\alpha\beta}(\textbf{r},t)=T^{\alpha\beta}_{nr}-\frac{v^{2}}{c^{2}}T^{\alpha\beta}_{nr}$$
$$+\frac{\hbar^{2}}{2mc^{2}}n(\partial^{\alpha}v^{\gamma}\partial^{\beta}v^{\gamma}+v^{\gamma}\partial^{\alpha}\partial^{\beta}v^{\gamma})
+\frac{\hbar^{2}}{4mc^{2}}n(v^{\alpha}\partial^{\beta}+v^{\beta}\partial^{\alpha})(\nabla \textbf{v})$$
$$+\frac{\hbar^{2}}{4mc^{2}}(\partial^{\gamma}n)\biggl(v^{\alpha}\partial^{\beta}v^{\gamma}+v^{\beta}\partial^{\alpha}v^{\gamma}\biggr)
-\frac{1}{c^{2}}\biggl(v^{\alpha}v^{\gamma}T^{\beta\gamma}_{nr}+v^{\beta}v^{\gamma}T^{\alpha\gamma}_{nr}\biggr)$$
$$-\frac{\hbar^{4}}{4m^{3}c^{2}}\biggl(\sqrt{n}\cdot\partial^{\alpha}\partial^{\beta}\triangle\sqrt{n}+\partial^{\alpha}\partial^{\beta}\sqrt{n}\cdot\triangle\sqrt{n}$$
\begin{equation}\label{SRW pressure simple}-\partial^{\alpha}\sqrt{n}\cdot\partial^{\beta}\triangle\sqrt{n}-\partial^{\beta}\sqrt{n}\cdot\partial^{\alpha}\triangle\sqrt{n}\biggr).\end{equation}
Relativistic part of this formula appears to contain two type of structures. Terms containing the velocity field $\textbf{v}$, so they explicitly have $v^{2}/c^{2}$ to reveal their semi-relativistic nature. The second type of structures is presented by the two last lines of formula (\ref{SRW pressure simple}). They have $\frac{\hbar^{2}}{mc^{2}}\triangle$ instead of $v^{2}/c^{2}$, so they give contribution in evolution of quantum plasmas even than velocity field $\textbf{v}$ is small. Transition from formula (\ref{Bohm potential}) to formula (\ref{SRW pressure simple}) is a step towards closing of the QHD set.

\subsection{Energy balance equation}

Now the energy balance equation (\ref{energy_eq}) should be expressed through the velocity field. After substitution of the Madelung decomposition (\ref{fla6}) into formula for the energy density and dividing it in two parts, we obtain
\begin{equation}
\label{fla8}
\varepsilon(\mathbf{r},t)=\varepsilon_{cl}(\mathbf{r},t)+\varepsilon_{q}(\mathbf{r},t),
\end{equation}
where $\varepsilon_{cl}$ is classical part of energy density (see (\ref{fla7})):
\begin{eqnarray}
&&\varepsilon_{cl}(\mathbf{r},t)=\int dR\sum_{i=1}^N\delta(\mathbf{r}-\mathbf{r}_i)a^2\biggl[\frac{1}{2}mv_i^2+\frac{3}{8c^2}mv_i^4
\nonumber\\
&&{}+\sum_{j=1,j\neq i}^N\biggl(\frac{1}{2}e^2G_{ij}+\frac{e^2}{4c^2}G^{\alpha\beta}_{ij}v_i^\alpha
v_j^\beta\biggr)\biggr],
\end{eqnarray}
and the quantum part of written below energy density $\varepsilon_{q}$ appears as:
\begin{eqnarray}
\label{fla9}
&&\varepsilon_{q}(\mathbf{r},t)=\frac{\hbar^2}{2m}\int dR\sum_{i=1}^N\delta(\mathbf{r}-\mathbf{r}_i)a^2\biggl[-
\frac{\Delta_i a}{a}+\frac{1}{4c^2}[(\nabla_i \textbf{v}_i)^2\nonumber\\
&&{}+2(\partial_i^\beta v_i^\gamma)(\partial_i^\beta v_i^\gamma)+2v_i^\beta\Delta_i v_i^\beta]+\frac{1}{c^2}\frac{\partial_i^\beta a}{a}(v_i^\beta\partial_i^\alpha v_i^\alpha+v_i^\alpha\partial_i^\beta v_i^\alpha)\nonumber\\
&&{}-\frac{1}{2c^2}a^{-1}\Delta_i a v_i^2+\frac{1}{c^2}a^{-2}\partial_i^\alpha a\partial_i^\beta a v_i^\alpha v_i^\beta-\frac{\hbar^2}{4m^2c^2}a^{-1}\Delta_i\Delta_i a\nonumber\\
&&{}+\sum_{j=1,j\neq i}^N\frac{e^2}{2mc^2}G^{\alpha\beta}_{ij}a^{-1}\partial_i^\alpha
\partial_j^\beta a\biggr].
\end{eqnarray}
First term in formula (\ref{fla9}) has non-relativistic quantum origin. Other terms are associated with semi-relativistic corrections, that is seen from presence of the factor $1/c^2$ before them.

Separating classical current (i. e. not proportional to the thermal velocities) and the Coulomb parts, function $\varepsilon(\mathbf{r},t)$ can be represented in following form:
\begin{eqnarray}
\varepsilon(\mathbf{r},t)&=&\frac{1}{2}mnv^2+\frac{3}{8c^2}mnv^4+\frac{e^2}{2}n\int d\mathbf{r}'G(\mathbf{r}-\mathbf{r}')n(\mathbf{r}',t)\nonumber\\
&&{}+\frac{e^2}{4c^2}nv^\alpha\int d\mathbf{r}'
G^{\alpha\beta}(\mathbf{r}-\mathbf{r}')n(\mathbf{r}',t)v^\beta(\mathbf{r}',t)+\rho\epsilon,
\end{eqnarray}
where $\rho\epsilon(\mathbf{r},t)$ may be called thermal energy density. This function is energy density without terms containing the velocity field only (i. e. without terms containing thermal velocities) in classical part of energy density $\varepsilon_{cl}$, and also the term, which is energy of the Coulomb interaction.

The energy current can be represented as follows:
\begin{eqnarray}
\label{fla10}
Q^\alpha=v^\alpha\varepsilon+v^\beta (p^{\alpha\beta}+T^{\alpha\beta})+q^\alpha.
\end{eqnarray}

The thermal current $q^\alpha$ entering in the last formula has
rather complicated form. So, we represent it as a sum of several
terms:
\begin{eqnarray}
\label{q as a sum of terms}
q^\alpha&=&\int dR\sum_{i=1}^N\delta(\mathbf{r}-\mathbf{r}_i)a^2\biggl[u_i^\alpha\tilde{\rho\epsilon}_i+u_i^\beta \tilde{T}^{\alpha\beta}_i\nonumber\\
&&{}-\frac{\hbar^2}{2m}a^{-1}\pib a\pia \vib-\frac{\hbar^2}{4m}\pia\pib\vib\biggr]\nonumber\\
&&{}+q^\alpha_0+q^\alpha_I+q^\alpha_{II}+q^\alpha_{III}+q^\alpha_{IV}+q^\alpha_{cur}.
\end{eqnarray}

Functions $\tilde{\rho\epsilon}_i$ and $\tilde{T}^{\alpha\beta}_i$ are introduced in the following way:
\begin{equation}\rho\epsilon=\int dR\sum_{i=1}^N\delta(\mathbf{r}-\mathbf{r}_i)\tilde{\rho\epsilon}_i,\end{equation}
\begin{equation}T^{\alpha\beta}=\int dR\sum_{i=1}^N\delta(\mathbf{r}-\mathbf{r}_i)\tilde{T}^{\alpha\beta}_i.
\end{equation}

Functions $q^\alpha_0$, $q^\alpha_I$, ..., $q^\alpha_{IV}$ are
associated with the order of derivative appearing before amplitude
of wave function $a(R,t)$; function $q^\alpha_{cur}$ is
contribution in the thermal current from the current-current
interaction. These functions are defined as
\begin{eqnarray}
q^\alpha_0&=&\int dR\sum_{i=1}^N\delta(\mathbf{r}-\mathbf{r}_i)\frac{\hbar^2}{4mc^2}a^2\biggl
[-\frac{1}{2}v_i^2\partial_i^\alpha\partial_i^\beta v_i^\beta\nonumber\\
&&{}+v_i^\gamma\partial_i^\alpha v_i^\beta\partial_i^\beta v_i^\gamma-v_i^\beta\partial_i^\alpha v_i^\gamma\partial_i^\beta v_i^\gamma
-v_i^\beta v_i^\gamma\partial_i^\alpha\partial_i^\beta v_i^\gamma\nonumber\\
&&{}+v_i^\alpha[(\partial_i^\beta v_i^\gamma)(\partial_i^\beta v_i^\gamma)+v_i^\beta \Delta_i v_i^\beta]-\frac{\hbar^2}{4m^2}\partial_i^\alpha\partial_i^\beta \Delta_i v_i^\beta\biggr],
\end{eqnarray}
\begin{eqnarray}
q^\alpha_I&=&\int dR\sum_{i=1}^N\delta(\mathbf{r}-\mathbf{r}_i)\frac{\hbar^2}{8mc^2}a\partial_i^\gamma a\biggl[-2v_i^2\partial_i^\alpha v_i^\gamma\nonumber\\
&&{}+4v_i^\alpha v_i^\beta\partial_i^\gamma v_i^\beta-\frac{\hbar^2}{m^2}\partial_i^\alpha\partial_i^\beta\partial_i^\gamma v_i^\beta-\frac{\hbar^2}{m^2}\partial_i^\alpha \Delta_i v_i^\gamma\biggr],
\end{eqnarray}
\begin{eqnarray}
q^\alpha_{II}&=&\int dR\sum_{i=1}^N\delta(\mathbf{r}-\mathbf{r}_i)\frac{\hbar^2}{4mc^2}\biggl
[-\frac{\hbar^2}{2m^2}a\Delta_i a\partial_i^\alpha\partial_i^\beta v_i^\beta\nonumber\\
&&{}-\frac{\hbar^2}{m^2}a\partial_i^\beta\partial_i^\gamma a \partial_i^\alpha\partial_i^\beta v_i^\gamma\nonumber\\
&&{}-a^2\partial_i^\alpha\partial_i^\beta\ln a\biggl\{3v_i^\beta v_i^2+\frac{\hbar^2}{2m^2}(\partial_i^\beta\partial_i^\gamma v_i^\gamma
+\Delta_i v_i^\beta)\biggr\}\biggr],
\end{eqnarray}
\begin{eqnarray}
q^\alpha_{III}&=&\int dR\sum_{i=1}^N\delta(\mathbf{r}-\mathbf{r}_i)\frac{\hbar^4}{4m^3c^2}[-a
\partial_i^\beta\Delta_i a \partial_i^\alpha v_i^\beta\nonumber\\
&&{}+(\partial_i^\alpha a\partial_i^\beta\partial_i^\gamma a-a\partial_i^\alpha\partial_i^\beta\partial_i^\gamma a)\partial_i^\beta v_i^\gamma],
\end{eqnarray}
\begin{eqnarray}
q^\alpha_{IV}&=&\int dR\sum_{i=1}^N\delta(\mathbf{r}-\mathbf{r}_i)\frac{\hbar^4}{8m^3c^2}(v_i^\alpha \Delta_i a\Delta_i a-2v_i^\beta a^{-1}\partial_i^\alpha a\partial_i^\beta a\Delta_i a\nonumber\\
&&{}+2v_i^\beta \Delta_i a\partial_i^\alpha\partial_i^\beta a-v_i^\alpha a\Delta_i\Delta_i a),
\end{eqnarray}
and
\begin{eqnarray}
\label{fla11}
&&q^\alpha_{cur}=\int dR\sum_{i,j=1,j\neq i}^N\delta(\mathbf{r}-\mathbf{r}_i)\frac{e^2\hbar^2a^2}{8m^2c^2}[G^{\alpha\beta}_{ij}(2\pjb v_{i\gamma}a^{-1}\pig a+\pjb\pig v_{i\gamma}\nonumber\\
&&{}+4a^{-1}\Delta_i a v_{j\beta})+G^{\beta\gamma}_{ij}(3\pia v_{j\gamma}a^{-1}\pib a+3\vjg\pia\pib\ln a\nonumber\\
&&{}+2\pia\pib\vjg+\pia v_{i\beta}a^{-1}\pjg a-\vib\pia\pjg\ln a)\nonumber\\
&&{}+\pia G^{\beta\gamma}_{ij}(\pib v_{j\gamma}+2\vjg a^{-1}\pib a)].
\end{eqnarray}

As a result, the energy balance equation take following form:
\begin{equation}
\label{fla15}
\partial_t \varepsilon(\mathbf{r},t)+\partial_\alpha Q^\alpha(\mathbf{r},t)=en(\mathbf{r},t)\mathbf{v}(\mathbf{r},t)\mathbf{E}+
\mathcal{A}(\mathbf{r},t),
\end{equation}
where functions $\varepsilon$ and $Q^\alpha$ are defined by formulas (\ref{fla8})-(\ref{fla9}) and (\ref{fla10})-(\ref{fla11}) correspondingly. Form of work density $\mathcal{A}$ is considered below in details. It is convenient to divide it on three parts:
$\mathcal{A}(\mathbf{r},t)=A^{cl}(\mathbf{r},t)+A^{q}(\mathbf{r},t)+\alpha(\mathbf{r},t)$. $A^{cl}(\mathbf{r},t)$ is part of work density corresponding to classical expression (see the right-hand side of equation (\ref{fla16})), $A^{q}(\mathbf{r},t)$ is quantum part of work density:
\begin{eqnarray}
&&A^{q}(\mathbf{r},t)=\frac{\hbar^2 e}{4m^2c^2}[\partial^\alpha(\partial^\beta E^\beta\cdot nv^\alpha)+\Delta E^\alpha nv^\alpha\nonumber\\
&&{}+\partial^\alpha E^\beta(v^\alpha\partial^\beta n+v^\beta\partial^\alpha n)+E^\alpha\partial^\alpha\partial^\beta(nv^\beta)]\\
&&{}-\frac{\hbar^2e^2}{8m^2c^2}\int d\mathbf{r}'\Delta G^{\alpha\beta}(\mathbf{r}-\mathbf{r}')[nv^\alpha(\mathbf{r},t)\partial_\beta^{'}n(\mathbf{r}',t)+
nv^\beta(\mathbf{r}',t)\partial_\alpha n(\mathbf{r},t)].\nonumber
\end{eqnarray}

Function $\alpha(\mathbf{r},t)$ has sense of thermal work density and can be represented as sum of three parts:
\begin{equation}
\alpha(\mathbf{r},t)=\alpha^{cl}(\mathbf{r},t)+\alpha^r(\mathbf{r},t)+\alpha^{cur}(\mathbf{r},t),
\end{equation}
where $\alpha^{cl}$ corresponds to classical part of thermal work density (\ref{fla12}),
\begin{eqnarray}
&&\alpha^r(\mathbf{r},t)=\int dR\sum_{i=1}^N\delta(\mathbf{r}-\mathbf{r}_i)\frac{e\hbar^2}{4m^2c^2}
[\partial_{i\alpha}(\partial_{i\beta}E_{i\beta}a^2u_{i\alpha})\nonumber\\
&&{}+\Delta_i E_{i\alpha}a^2u_{i\alpha}+E_{i\alpha}\partial_{i\alpha}\partial_{i\beta}(a^2u_{i\beta})
\nonumber\\
&&{}+2\partial_{i\alpha}E_{i\beta}(a\partial_{i\beta}a u_{i\alpha}+a\partial_{i\alpha}a u_{i\beta}+a^2\partial_{i\alpha}v_{i\beta})],
\end{eqnarray}
is completely quantum part of thermal work density associated with the relativistic correction to kinetic energy,
\begin{eqnarray}
&&\alpha^{cur}(\mathbf{r},t)=\int dR\sum_{i,j=1,j\neq i}^N\delta(\mathbf{r}-\mathbf{r}_i)\biggr[\frac{e^2\hbar^2}{8m^2c^2}
\partial_{i\gamma}G^{\alpha\beta}_{ij}\times\nonumber\\
&&{}\times\{v_{i\alpha}
a^2\partial_{i\gamma}\partial_{j\beta}\ln a+v_{j\beta}a^2\partial_{i\gamma}\partial_{i\alpha}\ln a\nonumber\\
&&{}-a(4v_{i\alpha}\partial_{i\gamma}\partial_{j\beta}a+
2v_{i\gamma}
\partial_{i\alpha}\partial_{j\beta}a+4v_{j\beta}
\partial_{i\gamma}\partial_{i\alpha}a\nonumber\\
&&{}+3\partial_{i\gamma}v_{i\alpha}\partial_{j\beta}a+
4\partial_{i\alpha}v_{j\beta}\partial_{i\gamma}
a+3\partial_{i\gamma}v_{j\beta}\partial_{i\alpha}a+
3\partial_{i\alpha}\partial_{i\gamma}v_{j\beta}a)\nonumber\\
&&{}-[v_{i\alpha}a^2\pjb\pjg\ln a+v_{j\beta}a^2\pia\pjg\ln a\nonumber\\
&&{}+2v_{j\gamma}a\pia\pjb a+\pjg v_{i\alpha}a\pjb a+\pjg v_{j\beta}a\pia a+\pjg\pia v_{j\beta}a^2]\}\nonumber\\
&&{}-\frac{e^2\hbar^2}{4m^2c^2}\Delta_i G_{ij}^{\alpha\beta}(a\pjb a u_{i\alpha}+a\pia a u_{j\beta}+a^2\pia v_{j\beta})\biggr],
\end{eqnarray}
is quantum part of thermal work density, presence of which is associated with the current-current interaction between particles. Function $\alpha^{cur}$, as the functions of thermal energy density $n\epsilon$, quantum Bohm potential $T_{\alpha\beta}$, thermal current $q^\alpha$, classical part of thermal work density $\alpha^{cl}$, involve velocities of particles, which are the sum of the velocity field and the thermal velocities. It should be noticed that these quantities involve velocity field, not thermal velocities only.

Equations (\ref{fla13}), (\ref{fla14}) and (\ref{fla15}) form a set of quantum hydrodynamical equations in the five-moment approximation including the continuity equation, the Euler equation and the energy balance equation.

It is necessary to notice that after introducing of the velocity field possibility to compare equations of quantum hydrodynamics with classics appears. These equations coincide to the accuracy of terms proportional to the Planck constant $\hbar$. In such a way possibility to derive quantum and classical equations at the same time. Classical equations obtain by turning to zero of Planck constant in the equations of quantum hydrodynamics.

In the absence of thermal contribution and many-particle quantum correlations we can present simplified form of quantum part of $\textbf{q}$ (\ref{q as a sum of terms}), which is an analog of the quantum Bohm potential (\ref{SRW pressure simple}) in the Euler equation (\ref{Euler quantum general form}). The quantum Bohm potential is the quantum part of the flux of particle current (the momentum current in the non-relativistic approach). Here we have the quantum part of the energy current presented as the sum of relativistic and non-relativistic parts:
\begin{equation}\label{q Bohm sum} \textbf{q}_{\hbar}=\textbf{q}_{non-rel\hbar}+\textbf{q}_{rel\hbar}.\end{equation}
The non-relativistic quantum energy current was obtained in Ref. \cite{LSK1999}. It can be written as
\begin{eqnarray}
q^\alpha_{non-rel\hbar}&=& -\frac{\hbar^2}{4m}(\partial_{\beta}n)\partial_{\alpha}v^{\beta}-\frac{\hbar^2}{4m}n\partial^{\alpha}(\nabla \textbf{v}).
\end{eqnarray}
The semi-relativistic part of the quantum energy current is rather large, so we present it as sum of several terms
\begin{eqnarray}
q^\alpha_{rel\hbar}&=& q^\alpha_{0,\hbar}+q^\alpha_{I,\hbar}+q^\alpha_{II,\hbar}+q^\alpha_{III,\hbar}+q^\alpha_{IV,\hbar},
\end{eqnarray}
where terms are separated by order of the spatial derivatives before the particle concentration.
These terms have following explicit form.
\begin{eqnarray}
q^\alpha_{0,\hbar}&=&\frac{\hbar^2}{4mc^2}n\biggl
[-\frac{1}{2}\textbf{v}^2\partial^\alpha(\nabla \textbf{v})\nonumber\\
&&{}+v^\gamma (\partial^\alpha v^\beta) \partial^\beta v^\gamma -v^\beta (\partial^\alpha v^\gamma) \partial^\beta v^\gamma
-v^\beta v^\gamma \partial^\alpha\partial^\beta v^\gamma\nonumber\\
&&{}+v^\alpha(\partial^\beta v^\gamma)(\partial^\beta v^\gamma)+ v^\alpha v^\beta \Delta v^\beta -\frac{\hbar^2}{4m^2}\partial^\alpha\partial^\beta \Delta v^\beta\biggr]
\end{eqnarray}
contains the particle concentration without derivatives of the particle concentration. The first six terms have the third order of the velocity field with two spatial derivatives. The last term has the first order of the velocity field, four spatial derivatives of the velocity field and additional square of the Plank constant.
\begin{eqnarray}
q^\alpha_{I,\hbar}&=& \frac{\hbar^2}{8mc^2}(\partial^\gamma n) \biggl[-\textbf{v}^2 \partial^\alpha v^\gamma\nonumber\\
&&{}+2v^\alpha v^\beta\partial^\gamma v^\beta -\frac{\hbar^2}{2m^2}\partial^\alpha\partial^\beta\partial^\gamma v^\beta
-\frac{\hbar^2}{2m^2}\partial^\alpha \Delta v^\gamma\biggr]
\end{eqnarray}
is proportional to gradient of the particle concentration. It also contains two group of terms. The first group has the third order of the velocity field with one spatial derivatives. The second group contains the first order of the velocity field with three spatial derivatives and additional square of the Plank constant.
\begin{eqnarray}
q^\alpha_{II,\hbar}&=& -\frac{\hbar^4}{4m^{3}c^2}\biggl
[\frac{1}{2}\sqrt{n}(\Delta \sqrt{n})\partial^\alpha\partial^\beta v^\beta+\sqrt{n}(\partial^\beta\partial^\gamma \sqrt{n}) \partial^\alpha\partial^\beta v^\gamma\nonumber\\
&&{}+\frac{1}{2}n(\partial^\alpha\partial^\beta\ln \sqrt{n})(\partial^\beta\partial^\gamma v^\gamma+\Delta v^\beta)\biggr]\nonumber\\
&&{}-\frac{\hbar^2}{4mc^2}3n(\partial^\alpha\partial^\beta\ln \sqrt{n})v^\beta \textbf{v}^2
\end{eqnarray}
consists of two groups of terms. The first of them consists of three terms. All of them contain the spatial derivatives of the concentration up to the second derivative. They also contain the second derivative of the velocity field. The second group consists of the one term. It has similar dependence on order of derivatives of the particle concentration as the first group of terms. It contains the third order of the velocity field without derivatives of the velocity field. It also explicitly shows $v^{2}/c^{2}$.
\begin{eqnarray}
q^\alpha_{III,\hbar}&=& \frac{\hbar^4}{4m^3c^2}[-\sqrt{n}
(\partial^\beta\Delta \sqrt{n}) \partial^\alpha v^\beta\nonumber\\
&&{}+((\partial^\alpha \sqrt{n})\partial^\beta\partial^\gamma \sqrt{n}-\sqrt{n}\partial^\alpha\partial^\beta\partial^\gamma \sqrt{n})\partial^\beta v^\gamma],
\end{eqnarray}
here we find the first spatial derivatives of the velocity field along with the spatial derivatives of the particle concentration. Each term may contain product of several concentrations, some of them may be under spatial derivatives, and total order of these derivatives is three.
\begin{eqnarray} \label{q Bohm 4}
q^\alpha_{IV,\hbar}&=& \frac{\hbar^4}{8m^3c^2}\biggl(v^\alpha (\Delta \sqrt{n})^{2}-\frac{1}{2}v^\beta \frac{1}{n\sqrt{n}}(\partial^\alpha n)(\partial^\beta n)(\Delta \sqrt{n})\nonumber\\
&&{}+2v^\beta (\Delta \sqrt{n}) \partial^\alpha\partial^\beta \sqrt{n}-v^\alpha \sqrt{n}\Delta\Delta \sqrt{n}\biggr)
\end{eqnarray}
contains the first order of the velocity field without spatial derivatives acting on it. $\textbf{q}_{IV,\hbar}$ has spatial derivatives of the particle concentration up to the fourth order.
Formulas (\ref{q Bohm sum})-(\ref{q Bohm 4}) is a step to get a closed set of the QHD equations in the five-moment approximation.

\section{Spectrum of linear excitations in semi-relativistic quantum plasmas}

Let us consider high-frequency linear excitations in semi-relativistic quantum plasmas consisting of electrons and ions (ions are assumed to be immobile). The number density and the velocity field are represented as $n=n_0+n'$, $v_\alpha=0+v'_\alpha$, $n_0$ is an equilibrium value of the number density; currents of particles are supposed to be absent in the equilibrium state. At linearization hydrodynamic equations take form
\begin{equation}\partial_t n'+n_0\partial_\alpha v'_\alpha=0,\end{equation}
$$m\partial_t v'_\alpha-\frac{\hbar^2}{4mn_0}\partial_\alpha\Delta n'-\frac{\hbar^4}{8m^3c^2n_0}\partial_\alpha\Delta\Delta n'+\gamma\frac{T}{n_0}\partial_\alpha n'=$$
$$=e\biggl[\partial_\alpha\varphi+\frac{\hbar^2}{4m^2c^2}
\partial_\alpha\Delta\varphi -\frac{5T}{2mc^{2}}\partial_\alpha\varphi-\frac{e^2n_0}{2mc^2}\int d\mathbf{r}'G_{\alpha\beta}(\mathbf{r}-\mathbf{r}')\partial_\beta^\prime\varphi(\mathbf{r}',t)
$$
\begin{equation}\label{linearized mom_bal_eq}
-\frac{e\hbar^2}{8m^2c^2}\int d\mathbf{r}'\partial_\gamma G_{\alpha\beta}(\mathbf{r}-\mathbf{r}')\partial_\beta^\prime\partial_\gamma^\prime n'(\mathbf{r}',t)\biggr],\end{equation}
and
\begin{equation}\Delta\varphi=4\pi en',\end{equation}
in the fourth term in the left-hand side of equation (\ref{linearized mom_bal_eq}) we use equation of adiabatic process and express pressure through temperature using equation of state $p=nT$ (where $T$ is temperature, $\gamma$ is the adiabatic index) and assuming that wave propagation is an adiabatic process.

Carrying Fourier transformation and taking into account that $G_{\alpha\beta}(\mathbf{k})=(8\pi/k^2)(\delta_{\alpha\beta}-
k_{\alpha}k_{\beta}/k^2)$, from condition of equality to zero for determinant of this set we obtain dispersion relation for the semi-relativistic Langmuir waves in plasma:
\begin{equation}\omega^2(k)=\omega_{p}^2\biggl(1-\frac{\hbar^2k^2}{4m^2c^2}-\frac{5T}{2mc^{2}}\biggr)+
\frac{\hbar^2k^4}{4m^2}-\frac{\hbar^4k^6}{8m^4c^2}+\frac{\gamma T}{m}k^2,\end{equation}where $\omega_p=\sqrt{4\pi e^2n_0/m}$ is the Langmuir frequency, $\gamma=3$ in classic adiabatic one-dimensional case (see \cite{Krall book}, chapter 4 section 4). This expression differs from classical one by presence of three terms, first of which is proportional to $k^2/c^2$ and it is consequence of new quantum semi-relativistic terms in the right-hand side of momentum balance equation. The first term proportional the Langmuir frequency contains three different contributions: the electric force giving the non-relativistic Langmuir oscillations, hybrid of the electric force and the semi-relativistic part of kinetic energy given by the last term of formula (\ref{rel correct to kin en in force field}) or the first term in formula (\ref{Force q with vel f}), and the thermal relativistic corrections having classic nature and presented by the first part of formula (\ref{rel correct to kin en in force field}) correspondingly. The second and third terms are proportional to $k^4$ and $k^6$ correspondingly and arise from formula obtained above for the quantum contribution in pressure given by the quantum Bohm potential (\ref{SRW pressure simple}).

In our recent paper \cite{Ivanov arxiv 12} we have considered contribution of the Darwin term in the Euler equation along with other semi-relativistic terms given by the Darwin Lagrangian and studied in this paper. We found that the hybrid force appearing at simultaneous account of the electric force and the semi-relativistic part of kinetic energy (see the first term in formula (\ref{Force q with vel f})) and the force field given by the Darwin interaction partially cancel each other living no trace in linear approximation. Consequently being considered together they do not give contribution in the spectrum of Langmuir waves. Therefore final semi-relativistic spectrum of the Langmuir waves is
\begin{equation}\label{Langmuir final}\omega^2(k)=\omega_{p}^2\biggl(1-\frac{5T}{2mc^{2}}\biggr)+
\frac{\hbar^2k^4}{4m^2}-\frac{\hbar^4k^6}{8m^4c^2}+\frac{\gamma T}{m}k^2.\end{equation}
This result also gives generalization of the spectrum obtained in Ref. \cite{Asenjo ZMBJ}, where authors did not considered contribution from the relativistic part of kinetic energy in equations of collective motion. They aimed to neglect contribution from relativistic part of the quantum Bohm potential (the third term in formula (\ref{Langmuir final})). However, it appears that the relativistic part of kinetic energy gives additional terms in the force field via commutators of corresponding terms in the Hamiltonian (\ref{Hamiltonian full as sum}) and the quantum particle current (\ref{fla3}).

\section{Conclusion}

During this work microscopic equations of quantum hydrodynamics for semi-relativistic system of particles based on the Darwin Hamiltonian. Equations were obtained in five-moment approximation including continuity equation, balance equation for momentum and energy. Role of quantum relativistic terms is emphasized in the momentum and energy balance equations. Unlike the Coulomb quantum plasma, where contribution of quantum effects is reduced mainly to the quantum Bohm potential and exchange interaction, in the considered case quantum relativistic effects lead to presence of terms with complex structure in the equations. In this paper we stay on derivation of equations and, as an illustration of quantum relativistic effects we consider dispersion of proper waves in semi-relativistic quantum plasma. From obtained dispersion relation it is seen that quantum relativistic effects appear even in linear approximation. So, we may expect appearing of various nonlinear quantum relativistic phenomena, at investigation of which equations derived in this paper can be applied.

\section{Appendix: Method of the Lorentz force extraction}

Let us consider the second term in the quantum force field describing the current-current interaction $\mathcal{F}^\alpha_{cur}$ (\ref{F curr quant}) since it gives the magnetic part of the Lorentz force. This term appears as
\begin{eqnarray}&&F_{L}^{\alpha}=\frac{e^{2}}{8m^{2} c^2}\int{dR}\sum_{i,j=1,i\neq j}^N\delta(\mathbf{r}-\mathbf{r}_i)(\partial_{i}^\alpha G^{\beta\gamma}_{ij}-
\partial_{i}^\beta G^{\alpha\gamma}_{ij})\times\nonumber\\
&&{}\label{L force tr 01}\times(\psi^*D_{i}^{\beta}D_j^\gamma\psi+
D_i^{*\beta}\psi^*D_j^\gamma\psi+c.~c.).\end{eqnarray}
We can represent it in terms of a two particle function
\begin{equation}\label{L force tr 02}F_{L}^{\alpha}=\frac{e^{2}}{2m^{2} c^2} \int (\partial^\alpha G^{\beta\gamma}(\textbf{r}-\textbf{r}')-
\partial^\beta G^{\alpha\gamma}(\textbf{r}-\textbf{r}'))\Im^{\beta\gamma}_{2}(\textbf{r},\textbf{r}',t) d\textbf{r}',\end{equation}
where
\begin{eqnarray}\label{L force tr 03} &&\Im^{\beta\gamma}_{2}(\textbf{r},\textbf{r}',t)=\int{dR}\sum_{i,j=1,i\neq j}^N\delta(\mathbf{r}-\mathbf{r}_i)\delta(\mathbf{r}'-\mathbf{r}_j)\times\nonumber\\ &&{}\times\frac{1}{4}(\psi^*D_{i}^{\beta}D_j^\gamma\psi+
D_i^{*\beta}\psi^*D_j^\gamma\psi+c.~c.).\end{eqnarray}
In the self-consistent field approximation a product of non-relativistic parts of the current $\textbf{j}$ (\ref{fla3}) arises in $\Im^{\alpha\beta}_{2}(\textbf{r},\textbf{r}',t)$. So we have $\Im^{\alpha\beta}_{2}(\textbf{r},\textbf{r}',t)=j^{\alpha}_{NR}(\textbf{r},t)
j^{\beta}_{NR}(\textbf{r}',t)$, where $\textbf{j}_{NR}(\textbf{r},t)$ is the non-relativistic part of current $\textbf{j}$ (\ref{fla3}). Since $\Im^{\alpha\beta}_{2}(\textbf{r},\textbf{r}',t)$ is in a semi-relativistic term we can present it as $\Im^{\alpha\beta}_{2}(\textbf{r},\textbf{r}',t)=j^{\alpha}(\textbf{r},t)
j^{\beta}(\textbf{r}',t)$, with the full currents $\textbf{j}$ (\ref{fla3}). Formula (\ref{L force tr 02}) now appears as
$$F_{L}^{\alpha}=\frac{e^{2}}{2m^{2} c^2} j^{\beta}(\textbf{r},t)\int (\partial^\alpha G^{\beta\gamma}(\textbf{r}-\textbf{r}')-
\partial^\beta G^{\alpha\gamma}(\textbf{r}-\textbf{r}'))j^{\gamma}(\textbf{r}',t)d\textbf{r}'$$
\begin{eqnarray}&&\label{L force tr 04}=\frac{e^{2}}{2m^{2} c^2} j^{\beta}(\textbf{r},t)\biggl[\partial^\alpha\int G^{\beta\gamma}(\textbf{r}-\textbf{r}')j^{\gamma}(\textbf{r}',t)d\textbf{r}'\nonumber\\
&&{}-\partial^\beta \int G^{\alpha\gamma}(\textbf{r}-\textbf{r}')j^{\gamma}(\textbf{r}',t)d\textbf{r}' \biggr].\end{eqnarray}

Introducing the vector potential of magnetic field created by moving charges
\begin{equation}\label{L force tr 05}A^{\alpha}_{int}(\textbf{r},t)=\frac{e}{2mc}\int G^{\alpha\beta}(\textbf{r}-\textbf{r}')j^{\beta}(\textbf{r}',t)d\textbf{r}',\end{equation}
and using the following identity
\begin{equation}\label{L force tr 06}\partial^\alpha A^\beta_{int}-\partial^\beta A^\alpha_{int}=\varepsilon^{\alpha\beta\gamma} B^{\gamma}_{int}\end{equation}
(it is meaning $\textbf{B}_{int}=\mathrm{curl}\textbf{A}_{int}$), we find
\begin{equation}\label{L force tr 07} F_{L}^{\alpha}=\frac{e}{mc}\varepsilon^{\alpha\beta\gamma}j^{\beta}B^{\gamma}_{int}.\end{equation}
Applying $\textbf{j}=mn\textbf{v}$ (see formula (\ref{velocity and j relation})) we can represent $\textbf{F}_{L}$ in the final form
\begin{equation}\label{L force tr 07}\textbf{F}_{L}=\frac{e}{c}n[\textbf{v}, \textbf{B}_{int}].\end{equation}
Let us admit that magnetic field $\textbf{B}_{int}$ obeys the magneto-static Maxwell equations $\mathrm{curl}\textbf{B}_{int}=\frac{4\pi}{c}\textbf{j}$ (\ref{QRD curlB}) and $\mathrm{div}\textbf{B}_{int}=0$ (\ref{QRD divB}).


\section{References}

\begin{thebibliography}{99}



\bibitem{LSK1999}L. S. Kuzmenkov, S. G. Maksimov, Theor. Math. Phys. \textbf{118}, N20, 227 (1999).

\bibitem{LSK2001} L. S. Kuzmenkov, S. G. Maksimov, V. V. Fedoseev, Theor. Math. Phys. \textbf{126}, N1, 110 (2001).


\bibitem{Andreev LSK Trukhanova PRB}P. A. Andreev, L. S. Kuzmenkov, M. I. Trukhanova, Phys. Rev. B \textbf{84}, 245401 (2011).


\bibitem{arxiv3335}P. A. Andreev, L. S. Kuz'menkov, Int. J. Mod. Phys. B \textbf{26}, N32, 1250186 (2012).

\bibitem{Andreev RPJ2007}P. A. Andreev, L. S. Kuz'menkov, Rus. J. Phys. \textbf{50}, N12, 1251 (2007).

\bibitem{LaLi2}L. D. Landau, E. M. Lifshitz, \emph{The Classical Theory of Fields} (Butterworth-Heinemann, 1975).

\bibitem{Vlasov UFN}A. A. Vlasov, Sov. Phys. Usp. \textbf{10}, 721 (1968).

\bibitem{Bohm Gross}D. Bohm, E. P. Gross, Phys. Rev. \textbf{75}, 1851 (1949).

\bibitem{LaLi10}L. P. Pitaevskii, E. M. Lifshitz, \emph{Physical Kinetics} (Pergamon Press, 1981).


\bibitem{Klimontovich book} Yu. L. Klimontovich, \emph{Statistical Physics} [in Russian], Nauka, Moscow (1982); English transl., Harwood, New
York (1986).

\bibitem{Zaslavskii 62} G. M. Zaslavskii, PMTF  N5, 42 (1962). (in Russian)

\bibitem{Pavlotskii DAN 73} I. P. Pavlotskii, DAN USSR \textbf{213}, N4, 812 (1973). (in Russian)

\bibitem{Pavlotskiy} Yu. N. Orlov, I. P. Pavlotsky, Matem. Mod. \textbf{1:12}, 31 (1989).

\bibitem{Pavlotsky P A 89 2}Yu. N. Orlov, I. P. Pavlotsky. Physica A \textbf{158}, 607 (1989).


\bibitem{Orlov MPL B 95}Yu. N. Orlov, V. V. Vedenyapin, Mod. Phys. Lett. B, \textbf{9}, N5, 291 (1995).



\bibitem{Orlov P A 92}Yu. N. Orlov, I. P. Pavlotsky, Physica A \textbf{184}, 558 (1992).


\bibitem{Mingalev PL A 96}O. V. Mingalev, Yu. N. Orlov, V. V. Vedenyapin, Phys. Lett. A \textbf{223}, 246 (1996).


\bibitem{Goldstein Ph 77} P. Goldstein, L. A. Turski, Physica \textbf{89A}, 481 (1977).

\bibitem{Jones 78 1} R. D. Jones, Phys. Fluids \textbf{21}, 2186 (1978).

\bibitem{Jones 78 2} R. D. Jones, Phys. Fluids \textbf{21}, 2191 (1978).

\bibitem{Jones PF 80 1} R. D. Jones, A. Pytte, Phys. Fluids \textbf{23}, 269 (1980).

\bibitem{Jones PF 80 2} R. D. Jones, A. Pytte, Phys. Fluids \textbf{23}, 273 (1980).



\bibitem{Wigner PR 84} M. Hillery, R.F. O'Connell, M.O. Scully, E.P. Wigner, Physics Reports, \textbf{106}, 121  (1984).



\bibitem{Shah PP 11} H. A. Shah, W. Masood, M. N. S. Qureshi, and N. L. Tsintsadze, Phys. Plasmas \textbf{18}, 102306 (2011).

\bibitem{Akbari-Moghanjoughi PP 13} M. Akbari-Moghanjoughi, Phys. Plasmas \textbf{20}, 042706 (2013).

\bibitem{Mahajan 03} S. M. Mahajan, Phys. Rev. Lett. \textbf{90}, 035001 (2003).

\bibitem{Asenjo PRE 12} F. A. Asenjo, F. A. Borotto, Abraham C.-L. Chian, V. Munoz,
J. A. Valdivia, E. L. Rempel, Phys. Rev. E \textbf{85}, 046406 (2012).

\bibitem{Lopez PRE 13} R. A. Lopez, F. A. Asenjo, V. Munoz, A. C.-L. Chian, and J. A. Valdivia, Phys. Rev. E \textbf{88}, 023105 (2013).

\bibitem{Ruyer PP 13} C. Ruyer, L. Gremillet, D. Benisti, and G. Bonnaud, Phys. Plasmas \textbf{20}, 112104 (2013).


\bibitem{Andreev arxiv rel} P. A. Andreev, arXiv:1208.0998.

\bibitem{Mahajan Yoshida}S. M. Mahajan, Z. Yoshida, Phys. Plasm. \textbf{18}, 055701 (2011).

\bibitem{Asenjo ZMBJ}F. A. Asenjo, J. Zamanian, M. Marklund, G. Brodin, P. Johansson, New J. Phys. \textbf{14}, 073042 (2012).


\bibitem{Andreev DSS 09} P. A. Andreev, L. S. Kuzmenkov, Dinamika
Slozhnykh Sistem  (2009) (in Russian).

\bibitem{pavelproc} P. A. Andreev and L. S. Kuzmenkov, PIERS Proceedings, Marrakesh, Morocco, March
20-23, p. 1047 (2011).

\bibitem{Trukhanova EPJD 13} M. I. Trukhanova,  Eur. Phys. J. D \textbf{67}, Issue 2, 32 (2013).

\bibitem{Trukhanova}M. I. Trukhanova, Int. J. Mod. Phys. B \textbf{26}, N1, 1250004 (2012).

\bibitem{Haas Eliasson Shukla PRE2012}F. Haas, B. Eliasson, P. K. Shukla, Phys. Rev. E \textbf{85}, 056411 (2012).

\bibitem{Asenjo Munoz PhysPlasm}F. A. Asenjo, V. Munoz, J. A. Valdivia, S. M. Mahajan, Phys. Plasm. \textbf{18}, 012107 (2011).


\bibitem{Uzdensky arxiv review 14} D. A. Uzdensky and S. Rightley, 	 arXiv:1401.5110.

\bibitem{Ivanov arxiv 12} A. Yu. Ivanov, P. A. Andreev, L. S. Kuz'menkov, arXiv: 1209.6124.

\bibitem{Ivanov RPJ 13} A. Yu. Ivanov and P. A. Andreev, Russ. Phys. J. \textbf{56}, 325 (2013).



\bibitem{Breit}G. Breit, Phys. Rev. \textbf{34}, 553 (1929).

\bibitem{LaLi4}V. B. Berestetskii, E. M. Lifshitz, L. P. Pitaevskii, \emph{Quantum Electrodynamics} (Butterworth-Heinemann, 1982).


\bibitem{Andreev LSK 2008}P. A. Andreev, L. S. Kuzmenkov, Phys. Rev. A \textbf{78}, 053624 (2008).


\bibitem{Haas}F. Haas. Phys. Plasm. \textbf{12}, 062117 (2005).

\bibitem{Marklund Brodin PRL}M. Marklund, G. Brodin, Phys. Rev. Lett. \textbf{98}, 025001 (2007).

\bibitem{Brodin Marklund 2007}G. Brodin, M. Marklund, New J. Phys. \textbf{9}, 277 (2007).



\bibitem{Madelung}E. Madelung, Z. Phys. \textbf{40}, 332 (1926).

\bibitem{Takabayashi}T. Takabayashi, Progr. Theor. Phys. \textbf{14}, 283 (1955).


\bibitem{Haas PRE 00} F. Haas, G. Manfredi, M. Feix, Phys. Rev. E \textbf{62},
2763(2000).


\bibitem{MaksimovTMP 2001 b} L. S. Kuz'menkov, S. G. Maksimov, and V. V. Fedoseev, Theor.
Math. Fiz. \textbf{126} 258 (2001) [Theoretical and Mathematical
Physics, \textbf{126} 212 (2001)].

\bibitem{Shukla Eliasson 2010 rus}P. K. Shukla, B. Eliasson, Phys. Usp. \textbf{53}, 51 (2010).

\bibitem{Shukla Eliasson 2011}P. K. Shukla, B. Eliasson, Rev. Mod. Phys. \textbf{83}, 885 (2011).


\bibitem{Kuzelev Ruhadze UFN1999}M. V. Kuzelev, A. A. Rukhadze, Phys. Usp. \textbf{42}, 603 (1999).

\bibitem{Andreev AtPhys 08} P. A. Andreev, L. S.  Kuz'menkov,
Physics of Atomic Nuclei \textbf{71},
N.10, 1724 (2008).

\bibitem{Andreev LSK arxiv12}P. A. Andreev, L. S. Kuz'menkov, Eur. Phys. J. D \textbf{67}, 216 (2013).



\bibitem{Andreev VestnMSU 2007} P. A. Andreev, L.S. Kuz'menkov,
Moscow University Physics Bulletin \textbf{62}, N.5, 271 (2007).

\bibitem{Brodin PRL 08} G. Brodin, M. Marklund, J. Zamanian, B. Ericsson and
P. L. Mana, Phys. Rev. Lett. \textbf{101}, 245002 (2008).

\bibitem{Vagin 09} D. V. Vagin, P. A. Polyakov, and N. E. Rusakova,
Moscow University Physics Bulletin, \textbf{64},  133
(2009).


\bibitem{Vladimirov PU 11} S. V. Vladimirov, Yu. O. Tyshetsky, Physics-Uspekhi \textbf{54}, 12 (2011).

\bibitem{Manfredi FIC 05} G. Manfredi, Fields Inst. Commun. \textbf{46}, 263 (2005).


\bibitem{Manfredi PRB 01} G. Manfredi and F. Haas, Phys. Rev. B \textbf{64}, 075316 (2001). 

\bibitem{Marklund TTSP 11} M. Marklund, J. Zamanian, G. Brodin, Transport Theory and Statistical Physics, \textbf{39}, 502
(2011).

\bibitem{Brodin PPCF 11} G. Brodin, M. Marklund, J. Zamanian and M. Stefan, Plasma Phys. Control. Fusion \textbf{53}, 074013 (2011).

\bibitem{Stefan PRE 11} M. Stefan, J. Zamanian, G. Brodin, A. P. Misra, and M. Marklund, Phys. Rev. E \textbf{83}, 036410 (2011).

\bibitem{Haas PL A 10} F. Haas, M. Marklund, G. Brodin, J. Zamanian, Phys. Lett. A \textbf{374}, 481 (2010).

\bibitem{Zamanian PP 10} J. Zamanian, M. Stefan, M. Marklund, and G. Brodin, Phys.Plasmas \textbf{17}, 102109 (2010).

\bibitem{Zamanian NJP 10} J. Zamanian, M. Marklund and G. Brodin, New J. Phys. \textbf{12}, 043019 (2010).

\bibitem{Haas NJP 10} F. Haas, J. Zamanian, M. Marklund and G. Brodin, New J. Phys. \textbf{12}, 073027 (2010).

\bibitem{Trovato JP A 10} M. Trovato, and L. Reggiani, J. Phys. A: Math. Theor. \textbf{43}, 102001 (2010).

\bibitem{Trovato PRE 10} M. Trovato, and L. Reggiani, Phys. Rev. E \textbf{81}, 021119 (2010).

\bibitem{Altaisky PL A 10} M. V. Altaisky, Phys. Lett. A, \textbf{374}, 522 (2010).

\bibitem{Maksimov TMP 02} L. S. Kuz'menkov and S. G. Maksimov, Theoretical and Mathematical Physics \textbf{131} 641 (2002).

\bibitem{Maksimov TMP 05} L. S. Kuz'menkov and S. G. Maksimov, Theoretical and Mathematical Physics \textbf{143}, 821 (2005).

\bibitem{Andreev kinetics 13} P. A. Andreev, arXiv:1308.3715.


\bibitem{Ivanov arxiv 13} A. Yu. Ivanov, L. S. Kuz'menkov, arXiv:1308.1966.

\bibitem{Koide PRC 13} T. Koide, Phys. Rev. C \textbf{87}, 034902 (2013).


\bibitem{Drofa1996}M. A. Drofa, L. S. Kuzmenkov, Theor. Math. Phys. \textbf{108}, N1, 849 (1996).

\bibitem{LSK 91}  L. S. Kuz'menkov, Theoretical and Mathematical
Physics \textbf{86}, 159 (1991).

\bibitem{pavelproc cl} P. A. Andreev and L. S. Kuzmenkov, PIERS Proceedings, Moscow, Russia, August 19-23, p. 158 (2012).


\bibitem{Landau Vol 3} L. D. Landau, E. M. Lifshitz, \emph{Quantum Mechanics: Non-Relativistic Theory}. Vol. 3 (3rd ed.). Pergamon Press,  (1977).

\bibitem{Krall book} N. A. Krall, A. V. Trivelpiece, \emph{Principles of Plasma Physics} (McGraw-Hill Book Company, 1973).
















\end{thebibliography}
\end{document}